\documentclass[aps,prb,twocolumn,superscriptaddress,footinbib,floatfix,showkeys,longbibliography]{revtex4-2}
\usepackage{graphicx}  % Include figure files
\usepackage{subfigure}
\usepackage{dcolumn} % Align table columns on decimal point
\usepackage{bm}% bold math
\usepackage{amssymb}   % for math
\usepackage{amsfonts}   % for math
\usepackage{comment}
\usepackage[colorlinks,linkcolor=blue,anchorcolor=blue,citecolor=blue,urlcolor=blue]{hyperref}
\usepackage{xcolor} 
\usepackage{soul}
\usepackage{tabularx,booktabs,ragged2e,multirow}
\newcolumntype{C}{>{\centering\arraybackslash}X}
\usepackage{multirow}
\usepackage{amsmath}
\usepackage{mathrsfs}
\usepackage{mathcomp}
\usepackage{textcomp}
\usepackage{dsfont}
\usepackage{esint}
\usepackage{braket}
\usepackage{textgreek}
\usepackage{lipsum}
\usepackage{marvosym} 
\usepackage{centernot}
\usepackage{cancel}
\usepackage{dashrule}
\usepackage{tikz}
\usepackage[compat=1.1.0]{tikz-feynman}
\def\shrinkage{0mu}
\def\vecsign{\mathchar"017E}
\def\dvecsign{\smash{\stackon[-1.95pt]{\mkern-\shrinkage\vecsign}{\rotatebox{180}{$\mkern-\shrinkage\vecsign$}}}}
\def\dvec#1{\def\useanchorwidth{T}\stackon[-4.2pt]{#1}{\,\dvecsign}}
\usepackage{stackengine}
\stackMath

\DeclareMathOperator{\Det}{Det}

\DeclareMathOperator{\sgn}{sgn}

\begin{document}
%\preprint{APS/123-QED}

\title{Topological spin textures in electronic non-Hermitian systems}

\author{Xiao-Xiao Zhang}
%\email{xiaoxiao.zhang@riken.jp}
\affiliation{RIKEN Center for Emergent Matter Science (CEMS), Wako, Saitama 351-0198, Japan}

\author{Naoto Nagaosa}
%\email{nagaosa@ap.t.u-tokyo.ac.jp}
\affiliation{Department of Applied Physics, University of Tokyo, Tokyo 113-8656, Japan}
\affiliation{RIKEN Center for Emergent Matter Science (CEMS), Wako, Saitama 351-0198, Japan}

%\date{\today}

\newcommand{\ba}{{\bm a}}
\newcommand{\bd}{{\bm d}}
\newcommand{\bb}{{\bm b}}
\newcommand{\bk}{{\bm k}}
\newcommand{\bmm}{{\bm m}}
\newcommand{\bn}{{\bm n}}
\newcommand{\br}{{\bm r}}
\newcommand{\bq}{{\bm q}}
\newcommand{\bp}{{\bm p}}
\newcommand{\bu}{{\bm u}}
\newcommand{\bv}{{\bm v}}
\newcommand{\bA}{{\bm A}}
\newcommand{\bB}{{\bm B}}
\newcommand{\bD}{{\bm D}}
\newcommand{\bE}{{\bm E}}
\newcommand{\bH}{{\bm H}}
\newcommand{\bJ}{{\bm J}}
\newcommand{\bK}{{\bm K}}
\newcommand{\bL}{{\bm L}}
\newcommand{\bM}{{\bm M}}
\newcommand{\bP}{{\bm P}}
\newcommand{\bR}{{\bm R}}
\newcommand{\bS}{{\bm S}}
\newcommand{\bX}{{\bm X}}
\newcommand{\brho}{{\bm \rho}}
\newcommand{\cA}{{\mathcal A}}
\newcommand{\cB}{{\mathcal B}}
\newcommand{\cG}{{\mathcal G}}
\newcommand{\cM}{{\mathcal M}}
\newcommand{\cP}{{\mathcal P}}
\newcommand{\bdelta}{{\bm \delta}}
\newcommand{\bgamma}{{\bm \gamma}}
\newcommand{\bGamma}{{\bm \Gamma}}
\newcommand{\bzero}{{\bm 0}}
\newcommand{\bOmega}{{\bm \Omega}}
\newcommand{\bsigma}{{\bm \sigma}}
\newcommand{\bUpsilon}{{\bm \Upsilon}}
\newcommand{\bcA}{{\bm {\mathcal A}}}
\newcommand{\bcB}{{\bm {\mathcal B}}}
\newcommand{\bcD}{{\bm {\mathcal D}}}
\newcommand\dd{\mathrm{d}}
\newcommand\ii{\mathrm{i}}
\newcommand\ee{\mathrm{e}}
\newcommand\zz{\mathtt{z}}
\newcommand\cE{\mathcal{E}}
\newcommand\cD{\mathcal{D}}
\newcommand\colonprod{\!:\!}

\makeatletter
\let\newtitle\@title
\let\newauthor\@author
\def\ExtendSymbol#1#2#3#4#5{\ext@arrow 0099{\arrowfill@#1#2#3}{#4}{#5}}
\newcommand\LongEqual[2][]{\ExtendSymbol{=}{=}{=}{#1}{#2}}
\newcommand\LongArrow[2][]{\ExtendSymbol{-}{-}{\rightarrow}{#1}{#2}}
\newcommand{\cev}[1]{\reflectbox{\ensuremath{\vec{\reflectbox{\ensuremath{#1}}}}}}
\newcommand{\red}[1]{\textcolor{red}{#1}} %for displaying red texts
\newcommand{\blue}[1]{\textcolor{blue}{#1}} %for displaying blue texts
\newcommand{\green}[1]{\textcolor{orange}{#1}} %for displaying blue texts
\newcommand{\mytitle}[1]{\textcolor{orange}{\textit{#1}}}
\newcommand{\mycomment}[1]{} %for commenting out
\newcommand{\note}[1]{ \textbf{\color{blue}#1}}
\newcommand{\warn}[1]{ \textbf{\color{red}#1}}

\makeatother

\begin{abstract}
Non-Hermitian systems have been discussed mostly in the context of open systems and nonequilibrium. Recent experimental progress is much from optical, cold-atomic, and classical platforms due to the vast tunability and clear identification of observables. However, their counterpart in solid-state electronic systems in equilibrium remains unmasked although highly desired, where a variety of materials are available, calculations are solidly founded, and accurate spectroscopic techniques can be applied. We demonstrate that, in the surface state of a topological insulator with spin-dependent relaxation due to magnetic impurities, highly nontrivial topological soliton spin textures appear in momentum space. Such spin-channel phenomena are delicately related to the type of non-Hermiticity and correctly reveal the most robust non-Hermitian features detectable spectroscopically.
Moreover, the distinct topological soliton objects can be deformed to each other, mediated by topological transitions driven by
tuning across a critical direction of doped magnetism. 
These results not only open a solid-state avenue to exotic spin patterns via spin- and angle-resolved photoemission spectroscopy, but also inspire non-Hermitian dissipation engineering of spins in solids.
\end{abstract}
%\pacs{71.10.Pm, 71.27.+a, 72.15.Nj, 72.15.Rn}
\keywords{non-Hermitian system, topological spin texture, magnetic impurity, surface state, ARPES}

\maketitle
% \tableofcontents
% \newpage
% \clearpage

%\clearpage
\let\oldaddcontentsline\addcontentsline% Store \addcontentsline
\renewcommand{\addcontentsline}[3]{}% Make \addcontentsline a no-op

\section{Introduction}
The physical properties of open and nonequilibrium systems are usually described 
as non-Hermitian (NH) systems in sharp contrast to that of closed systems governed by the Hermitian Hamiltonian and corresponding unitary operator. NH systems show a variety of unique properties compared to Hermitian 
ones, which have been demonstrated both theoretically and experimentally \cite{Bender2007,Mostafazadeh2010,Guo2009,*Rueter2010,*Schindler2011,Moiseyev2009,Torres2019,Ashida2020, Bergholtz2021}. 
For instance, complex eigenenergies can be degenerate in the form of exceptional points (EPs) of the NH Hamiltonian, where eigenstates coalesce into one and highly enhanced sensitivity and many-body phase transitions can occur \cite{Berry2004,Heiss2012,Dembowski2004,Gao2015,Zhen2015,Wiersig2014, Chen2017,Hodaei2017,Su2021,Zhang2021b,Fruchart2021,Bergholtz2021}. 
Topology of NH systems has also been revealed to differ from conventional topological systems \cite{Shen2018, Gong2018,Torres2019,Kawabata2019a,Ashida2020}. 
Another interesting phenomenon is the NH skin effect, where many states are localized near the boundary, leading to the breakdown of conventional bulk-boundary correspondence \cite{Alvarez2018a,Kunst2018,Yao2018a,Xiong2018,Lee2019,Yokomizo2019, Helbig2020,Xiao2020,Ghatak2020,Okuma2020,Zhang2020,Lu2021,Zhang2022a,Zhang2023}.

Since quasiparticles are subject to the relaxation from coupling to other degrees of freedom, non-Hermiticity should also be a common feature in many-body quantum systems even in equilibrium. Notwithstanding, the existing experiments aforementioned are mostly from optical, cold-atomic, and classical platforms; no essential experimental progress has appeared in solid-state electronic systems although it is much longed for from a condensed matter viewpoint. Such a very unusual and unsatisfactory situation lies in that, in terms of NH properties, solid-state electrons in equilibrium have fundamental differences. Other NH platforms, typically bosonic, purely spin, or classical, are often not charged fermions and are able to be pushed toward a controllable nonequilibrium state: this will strongly affect the nature of physical processes involved in detections and experiments. % such as photoemission spectroscopy. 
While solid-state systems are rich in platforms, calculation methods, and detection means, the main obstacle is that it remains largely unclear what the robust and distinguishing \textit{observables} should be. It is therefore an urgent need to clarify and find the true observable effects of solid-state NH electronic systems.
For instance, although exceptional degeneracies are proven to be crucial in optical systems, do they fundamentally affect solid-state electronic spectroscopy? If not, what to observe?
Here, we address these related and pressing issues in a highly experimentally relevant way and point out a concrete solution. Directly aiming at the powerful spin- and angle-resolved 
photoemission spectroscopy (SARPES) \cite{Lv2019,Sobota2021}, we predict exotic spin textures including monopoles, skyrmions, vortices, and merons. They reveal the topologically robust spectroscopic features to be detected for realistic NH systems, overlooked in previous perspective of NH physics and representing a new route towards solid-state electronic dissipation engineering of spins.

In fact, impurity scattering, 
electron-phonon interaction, and electron-electron interaction can lead to the 
self-energy %of electronic Green's function 
with its imaginary part ${\rm Im}\Sigma$ accommodating a NH perspective \cite{Kozii2017, Zyuzin2018, Yamamoto2019, Zyuzin2019}. 
The NH nature becomes particularly evident when ${\rm Im}\Sigma$ gains internal structure, e.g., orbital or band dependence.
Exceptional degeneracies and related bulk Fermi arcs in Kondo and other correlated systems is one example \cite{Yoshida2018,Nagai2020,Michishita2020}.
With magnetism and/or spin-orbit interaction (SOI), it is generally expected that ${\rm Im}\Sigma$ depends on spin.
For instance, from the matrix element of magnetic impurity scattering $U_{k\sigma, k'\sigma'} = u_0 + \sigma u \delta_{\sigma, \sigma'}$, one can find that ${\rm Im}\Sigma_\sigma = \gamma_0 + \sigma \gamma$
with $\gamma_0 \propto u_0^2 + u^2$ and $\gamma \propto 2u_0 u$ [see Supplemental Material (SM) \ref{SecSM:imp}].
Two damping constants $\gamma_0$ and $\gamma$ satisfies $\gamma_0 > \gamma$, representing the stability of the system
in equilibrium. 
Relativistic SOI often crucially determines the momentum-space spin texture as
been demonstrated in Rashba systems and the surface state of three-dimensional topological insulators \cite{Topo1,*Topo2, Hsieh2009, Xu2016}. Spin splitting and 
spin-momentum locking occur in these noncentrosymmetric systems; relaxation is often regarded to play a minor role at on-shell region with large spin splitting. However, the NH spin-dependent relaxation can have significant and robust influence on the spin texture, which we address below in the surface state context.

\section{Models and methods}
\subsection{System with effective NH description} 
We focus on the two-dimensional Dirac model
\begin{equation}\label{eq:H0}
    H_0=d_\nu(\bk)\sigma^\nu=v(k_x\sigma_2 - \chi k_y\sigma_1) + m\sigma_3,
\end{equation}
which represents the (magnetic) surface state with spin matrices $\sigma_\nu=(\sigma_0,\bsigma)=(I,\sigma_1,\sigma_2,\sigma_3)$. Usually $\chi=1$ while  $\chi=-1$ represents the opposite intrinsic chirality possible when rotational symmetry such as $C_{n>2}$ is broken. Although the mass term is here fixed to $\sigma_3$, the $\gamma_1$-relaxation below is  often associated with doped magnetization along $\hat{x}$, which can be readily related as explained later. We henceforth set physical constants $e,\hbar,k_\mathrm{B}$ and the Fermi velocity $v$ to unity and do not distinguish the 4-vector subscript or superscript.
$d_0=0$ will also be used in numerics for its irrelevance to the spin channel.
To generally account for the relaxation, one can attach a spinful bath in the wide-band limit, leading to the self-energies in the Keldysh formalism $    \Sigma_B^{\mathrm{r}(\mathrm{a})} = \mp\ii\gamma_\nu\sigma^\nu,
\Sigma_B^< =  2\ii\gamma_\nu f(\omega)\sigma^\nu$ with $f(\omega)$ the Fermi distribution, detailed in SM \ref{SecSM:photoemission} \cite{Rammer2011,Stefanucci2015}. %They will also be used for the spectroscopy calculation.  

In the following, we focus on two representative cases, i) $\bgamma=\gamma_3\hat{z}$ and ii) $\bgamma=\gamma_1\hat{x}$, and then discuss general $\bgamma$. We have $\gamma_0>\gamma=|\bgamma|$ and assume $\gamma_\nu>0$ for concreteness. The NH effective Hamiltonian is given by  
\begin{equation}\label{eq:H_eff}
    H_\mathrm{eff}=H_0+ H_1
\end{equation}
with the NH part 
\begin{equation}
    H_1=\Sigma_B^\mathrm{r}=-\ii\gamma_\nu\sigma^\nu,
\end{equation}
which is the starting point of most studies on NH physics.
For the massless case of i), $H_\mathrm{eff}$ exhibits an exceptional ring (ER) with radius $\gamma_3$ and centered at $\bk=0$, inside which lies the bulk Fermi drumhead state.
When $\gamma_1> |m|$, $H_\mathrm{eff}$ of case ii) exhibits two EPs located at $(\pm\sqrt{\gamma_1^2-m^2},0)$ and connected by the bulk Fermi arc along $k_x$-axis where the real parts of the two bands touch. In fact, NH systems represented by $H_\mathrm{eff}$ are topologically nontrivial and characterized by $\mathbb{Z}$, especially the more general gapped case without EPs bears a NH Chern number (SM \ref{SecSM:NHtopo}) \cite{Shen2018,Okugawa2019,*Yoshida2019,Kawabata2019a,*Kawabata2019b}. 
We will see how the spectroscopically observable spin topology shows distinct and richer structures than such classification while maintaining some connection.

\subsection{Spin-resolved photoemission spectroscopy}\label{Sec:spectroscopy}

For solid-state electrons with interaction/disorder induced NH relaxation, although effective Hamiltonians contain important information about the system, they may not be the appropriate choice to describe some experimental phenomena.
This is at the root of the hindered experimental progress aforementioned.
For instance, simply solving the effective Hamiltonian Eq.~\eqref{eq:H_eff}, its complex energy spectrum can apparently blur the identification of ground state spin texture, and the possible exceptional degeneracy makes one spin state ill-defined. These issues from the effective Hamiltonian viewpoint are not directly related to realistic spin-resolved spectroscopy, which can always be unambiguously performed with any NH relaxation present. Instead, as a merit of the solid-state electronic system, we propose the more complete Green’s function formalism in order to properly and fully incorporate these NH relaxation effects, where the same NH information is included as in the effective Hamiltonian.

We apply the SARPES formalism \cite{Freericks2009,Zhang2022} and obtain the density and spin-resolved polarization $P_\nu(\omega,\bk) = 1/2\pi \int\dd \omega' \rho_\nu(\omega',\bk)\ee^{-(\omega-\omega')^2t_\mathrm{p}^2}$
% \begin{equation}\label{eq:P_nu}
% \begin{split}
%     P_\nu(\bk,\omega) &= \mathrm{Tr}[\sigma_\nu\, \frac{-\ii}{2\pi}\int\dd \omega' \cG^<(\bk,\omega')\ee^{-(\omega-\omega')^2t_\mathrm{p}^2}] .
% \end{split}
% \end{equation}
with the expectation value %(see \textit{Methods})
\begin{equation}\label{eq:rho_nu}
    \rho_\nu=\mathrm{Tr}[-\ii\sigma_\nu\cG^<]  =  4 f(\omega)p_\nu/|\tilde{\varepsilon}^2-\tilde{d}^2|^2. 
\end{equation}
We denote $\tilde{\varepsilon}=\varepsilon+\ii\varepsilon'$, $\tilde{\bd}=\bd+\ii\bd'$,  $\varepsilon=\omega-d_0,\varepsilon'=\gamma_0,\bd'=-\bgamma$. 
When the experimentally controllable probe pulse width $t_\mathrm{p}$ is wide, one can directly focus on the observable $\rho_\nu$ from lesser Green's function $\cG^<=\cG^\mathrm{r} \Sigma_B^< \cG^\mathrm{a}$ in equilibrium states, where $\cG^\mathrm{r(a)}=(\omega-H_0-\Sigma_B^\mathrm{r(a)})^{-1}$. 
Therefore, the photoemission spectroscopy involves all the aforementioned self-energies and one finds %$\rho_\nu=\mathrm{Tr}[-\ii\sigma_\nu\cG^<]  =  \frac{4 f(\omega)}{|\tilde{\varepsilon}^2-\tilde{d}^2|^2} p_\nu$,
% \begin{equation}\label{eq:Trg^<}
% \begin{split}
%     \mathrm{Tr}[\sigma_\nu\,\cG^<] & =  \frac{4\ii f(\omega)}{|\tilde{\varepsilon}^2-\tilde{d}^2|^2} p_\nu ,
% \end{split}
% \end{equation}
the real 4-vector $p_\nu$ in Eq.~\eqref{eq:rho_nu} 
\begin{equation}\label{eq:p_nu_maintext}
\begin{split}
    p_0& = \gamma_0(\varepsilon^2+E^2) + 2 \varepsilon\bd\cdot\bgamma \\
    \bp &= 2(\gamma_0\varepsilon+\bd\cdot\bgamma)\bd + (\varepsilon^2-E^2) \bgamma
\end{split}
\end{equation}
with $E^2=d^2+\gamma_0^2-\gamma^2$. Although the positive-definite density channel $\rho_0(\varepsilon,\bk)$ can reflect some overall symmetry of the relaxation $\bgamma$, there is \textit{no} distinguishing feature beyond merely quantitative variation in the profile. SM \ref{SecSM:density_channel} provides concrete examples of the irrelevance of exceptional degeneracies, which is based on the Hamiltonian viewpoint, in photoemission spectroscopy. The inadequacy of a straightforward Hamiltonian viewpoint has also been noticed for electronic systems in terms of the NH skin effect \cite{Okuma2021,Lee2020}. This partially explains the lack of experimental progress as aforementioned. 
While exceptional degeneracies from the Hamiltonian approach are neither the essential physics nor unambiguously identifiable in the present experimental detection, $\bp(\varepsilon,\bk)$ defined above in the energy-momentum space represents the physically most informative quantity with new observable features, especially because it fully determines the topology of the momentum-space spin texture. Note that it is based on a physically more complete description of the same NH electronic system.
We henceforth mainly focus on $\bp$ instead of $p_0$ or $\rho_0$. The temperature and occupation affect the global strength of spectroscopic signals following the Fermi distribution $f(\omega)$ in Eq.~\eqref{eq:rho_nu}.

It is as well worth noting that finite relaxation rate $\gamma_\nu$ in general broadens $\cG^<$ and hence the photoemission signal in off-shell regions. For instance, near $\varepsilon=0$, i.e., $\omega\approx d_0$, it is not spectroscopically empty even with a finite gap $m\neq0$.
For the simpler case with only $\gamma_0$ present, Eq.~\eqref{eq:p_nu_maintext} reduces to $p_0=\gamma_0(\varepsilon^2+E^2),\bp= 2\gamma_0\varepsilon\bd$
% \begin{equation}\label{eq:p_gamma0}
% \begin{split}
%     p_0=\gamma_0(\varepsilon^2+E^2),\quad\bp= 2\gamma_0\varepsilon\bd
% \end{split}
% \end{equation}
where the original surface-state spin-momentum locking manifests in $\bp$.  
When $\gamma_0\rightarrow0$, $\cG^<$ will attain $\delta$-function peaks for the on-shell energies and momenta, representing the original band structure with the typical spin-momentum locking texture (see SM \ref{SM:nogamma} and SM Fig.~\ref{Fig:hermitian}).
Here, we note again that, for most experimentally accessible solid-state systems without external pumping, the overall effect would be dissipation or loss rather than gain, i.e., $\gamma_0>\gamma$. This will have immediate consequences on all the observable effects, while the scenario under light irradiation will be studied elsewhere.
However, even under this constraint, once the band-dependent NH effect or spinful relaxation is taken into account, interesting spin structures can emerge.

\section{Results}
\subsection{Skyrmion and monopole textures from $\gamma_3$-relaxation}

\begin{figure*}[hbt]
\includegraphics[width=17.8cm]{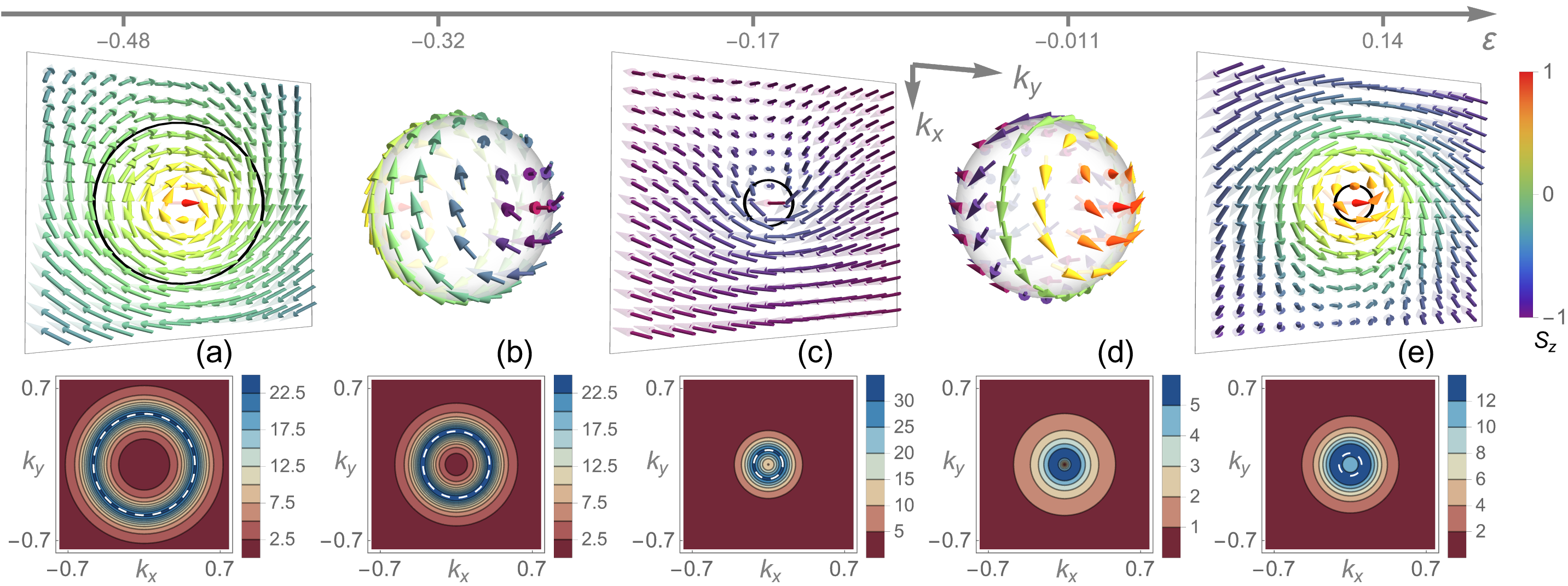}
\caption{ (Top) Topological spin textures in the $(\varepsilon,\bk)$-space with $\gamma_3$-relaxation. (a,c,e) Three energy planes in the square-shaped $\bk$-space $[-0.8,0.8]^2$. (b,d) A pair of (b) antimonopole and (d) monopole are respectively centered at $(\varepsilon^\mp,\bzero)$; spin texture on the spheres of radius $0.2$ surrounding them are plotted. Energy plane (a) lower than antimonopole $\varepsilon^-$ or (e) higher than monopole $\varepsilon^+$ exhibits a nontrivial Bloch-type skyrmion while (c) in between the monopole pair shows an in-plane vortex only. Black circle indicates the on-shell ring on the Dirac cone cut by $\varepsilon$-plane.  
(Bottom) Intensity contour profile of the spin channel signal $|\brho(\varepsilon,\bk)|$ at zero temperature is in one-to-one correspondence with the top panel and the on-shell ring is now indicated by a white dashed circle. (b,d) correspond to the energy planes $\varepsilon^\mp$ crossing the monopole cores.
Parameters are $\chi=1,m=0.1,\gamma_0=0.1,\gamma_3=0.06,\mu=0.5$.
}\label{Fig:skyrmion}
\end{figure*}

When $\bgamma=\gamma_3\hat{z}$,
we have the in-plane and out-of-plane part of $\bp$ from Eq.~\eqref{eq:p_nu_maintext}
\begin{equation}\label{eq:p_nu_gamma3_maintext}
\begin{split}
    \bp_{12}&=(p_1,p_2) =2(\gamma_0\varepsilon+m\gamma_3)\bd_{12} \\
    p_{3}&=\gamma_3[\varepsilon(\varepsilon+2m\gamma_0/\gamma_3)+m^2-(\gamma_0^2-\gamma_3^2)-k^2]
\end{split}
\end{equation}
where $\bd_{12}=(d_1,d_2)$.
As shown in Fig.~\ref{Fig:skyrmion}, this $\bk$-space spin texture $\bS=\bp/|\bp|$ consists of a Bloch-type \textit{skyrmion} \cite{Sk:review1,Goebel2021} centered at $\bk=\bzero$ as long as $\varepsilon(\varepsilon+2m\frac{\gamma_0}{\gamma_3})+m^2>\gamma_0^2-\gamma_3^2$, i.e., when either $|m|$ or $|\varepsilon|$ is large enough to overcome the difference between $\gamma_0$ and $\gamma_3$. This is entirely absent in the Hermitian surface state (SM Fig.~\ref{Fig:hermitian}). Here, for the simple situation $\varepsilon=0$ ($m=0$), the condition to form a skyrmion is $m^2>\gamma_0^2-\gamma_3^2$ ($\varepsilon^2>\gamma_0^2-\gamma_3^2$): close to the band midpoint $\varepsilon\approx0$ finite mass is necessary; otherwise, e.g., for typically occupied $\varepsilon<0$ branch $\varepsilon$ itself suffices even when massless. 
This condition, which is made possible purely by the presence of finite $\gamma_3$, guarantees a sign change of the out-of-plane component $p_3$. Together with the in-plane winding of $\bp_{12}$ inherited from SOI, as momentum $k$ increases spins pointing towards the northern hemisphere eventually change to the southern hemisphere and a skyrmion is realized.

The emergence of skyrmionic structure becomes even more complete and richer if viewing in the three-dimensional $(\varepsilon,\bk)$-space. There actually exist two singular points along the $\bk=\bzero$ line at %$\varepsilon^s=[-m\gamma_0+s(\gamma_0^2-\gamma_3^2)^\frac{1}{2}(m^2+\gamma_3^2)^\frac{1}{2}]/\gamma_3$
\begin{equation}
    \varepsilon_\mathrm{mp}^s=[-m\gamma_0+s(\gamma_0^2-\gamma_3^2)^\frac{1}{2}(m^2+\gamma_3^2)^\frac{1}{2}]/\gamma_3
\end{equation}
for $s=\pm1$ and none elsewhere where the spin orientation is ill-defined. Shown in Figs.~\ref{Fig:skyrmion}(b,d), they form a \textit{monopole-antimonopole pair} that germinates skyrmions. When $\chi=1$ ($\chi=-1$) energy planes in between the monopole (antimonopole) at $\varepsilon_\mathrm{mp}^+$ and the antimonopole (monopole) at $\varepsilon_\mathrm{mp}^-$, e.g., Fig.~\ref{Fig:skyrmion}(c), do not exhibit skyrmion but vortex, while those outside in Figs.~\ref{Fig:skyrmion}(a,e) do. 
As physically expected, the hedgehog-type spin textures near the two cores take the form linear in $\bdelta\equiv(\delta\varepsilon,\delta\bk)$ measured from the cores
% $\delta\bp^\pm=2(\gamma_0\varepsilon^\pm+\gamma_3m)\delta\bd_{12} + 2(\gamma_0m+\gamma_3\varepsilon^\pm)\delta\varepsilon\hat{z}$
\begin{equation}\label{eq:deltap^pm_main}
\begin{split}
    \delta\bp^\pm=2(\gamma_0\varepsilon^\pm+\gamma_3m)\delta\bd_{12} + 2(\gamma_0m+\gamma_3\varepsilon^\pm)\delta\varepsilon\hat{z}  
\end{split}
\end{equation}
where $\delta\bd_{12}=(-\chi\delta k_y,\delta k_x)$. 
To understand the formation of NH topological structures, we note that the NH relaxation is spin dependent and selects a direction in spin space, which is combined with the spin texture from the spin-momentum locking. More explicitly, 
the monopole and skyrmion can inherit the in-plane winding feature, as reflected by the vorticity in Eq.~\eqref{eq:N_sk_gamma3} below, from the original surface state; besides, tiny NH relaxations create one monopole in the pair near $\varepsilon=0$ and the other from energy infinity for the generic massive case, which further moves inside with increasing $\gamma_3$ (SM \ref{SecSM:spin_channel}). 
The skyrmion is thus a natural consequence since its quantized topological flux at each energy plane is directly associated with and from the monopole charge, which is a picture resembling a minimal Weyl semimetal with two Weyl points of opposite charges and associated quantum anomalous Hall states in momentum slices \cite{Weyl2011,WeylDiracReview}. See also Sec.~\ref{Sec:summary} for further discussion of the formation mechanism.

Parametrizing the spin orientation defined in the polar coordinates of the momentum space $\bS(k,\phi)=(\sin\Theta\cos\Phi,\sin\Theta\sin\Phi,\cos\Theta)$, a skyrmion is characterized by its polarity $\mathtt{p}=-\frac{1}{2}\cos\Theta(k)\big|_{0}^\infty$, vorticity $\mathtt{v}=\frac{1}{2\pi}\Phi(\phi)\big|_0^{2\pi}$, and helicity $\mathtt{h}$ from $\Phi=\mathtt{v}\phi+\mathtt{h}$; the skyrmion number is $N_\mathrm{sk}=\mathtt{p}\mathtt{v}$.
The present case bears 
\begin{equation}\label{eq:N_sk_gamma3}
    N_\mathrm{sk}=\chi,\mathtt{p}=1,\mathtt{v}=\chi,\mathtt{h}=\pm\pi/2
\end{equation}
and the sign of helicity is given by $\sgn[(\gamma_0\varepsilon+m\gamma_3)\chi]$. For the massless case, which has an aforementioned NH effective ER, its radius $\gamma_3$ participates in partially controlling the size of the skyrmion or where the spins point in-plane. However, the emergence of skyrmion/monopole is not limited to the case where ER exists. It is a distinct and even stabler feature of the NH $\gamma_3$ term beyond the effective Hamiltonian $H_\mathrm{eff}$ itself, which also holds in other cases below. This is consistent with our earlier discussion and expectation that the new formalism will surpass the exceptional degeneracy.
Lastly, we mention that, for the large-$(\varepsilon,\bk)$ region close to the band dispersion, the texture will eventually reproduce the case with only $\gamma_0$. Such a crossover mainly happens in the out-of-plane channel, where the pointing will be fixed by the on-shell condition and lose the sign switching associated with increasing $k$.

In the bottom panel of Fig.~\ref{Fig:skyrmion}, we provide the information of the SARPES signal strength of the spin channel $\rho(\omega,\bk)=|\brho(\omega,\bk)|$ of Eq.~\eqref{eq:rho_nu}, in addition to the topological features of the normalized spin textures highlighted in the top panel.
We see that the intensity is circularly symmetric as $\gamma_3$ does not pick an in-plane direction. The on-shell ring or annulus region is the strongest in Fig.~\ref{Fig:skyrmion}(a,b,c) as the energy plane moves upward until the gap region starts at $\varepsilon=-m$, which is consistent with the geometry of the lower branch of the Dirac cone. As the gap is traversed and the energy gradually moves to the upper branch, the signal in Fig.~\ref{Fig:skyrmion}(d) is mainly contributed by the relaxation broadening and hence relatively small while that in Fig.~\ref{Fig:skyrmion}(e) once again gets large since it is below the chemical potential. The asymmetry in the signal strength with respect to $\varepsilon=0$ follows the asymmetric position of the monopole pair, which owes to the chiral symmetry breaking due to finite mass and relaxation $\gamma_3$ \cite{Okugawa2019,*Yoshida2019}. 
At large momenta, the signal will decay as it becomes further and further away from the on-shell region, although there is residue broadening due to the relaxation effects.

\subsection{Meron and vortex pair textures from $\gamma_1$-relaxation}

\begin{figure*}[hbt]
\includegraphics[width=17.8cm]{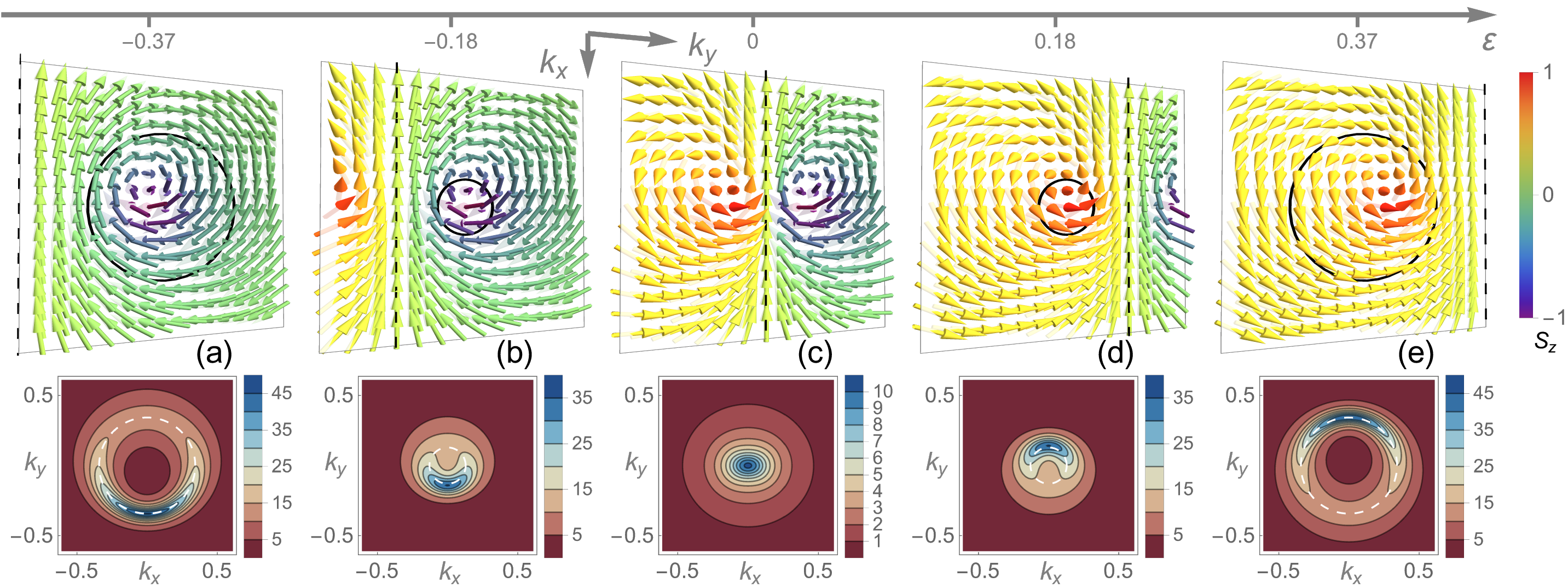}
\caption{(Top) Topological spin textures in the $(\varepsilon,\bk)$-space with $\gamma_1$-relaxation. (a,b,c,d,e) Five energy planes in the square-shaped $\bk$-space $[-0.6,0.6]^2$. A meron-antimeron pair appears in each energy plane. Dashed black line indicates the borderline $k_y=K_y^0(\varepsilon)$ between the meron and antimeron respectively centered at $\bK^\mp(\varepsilon)$. As the borderline shifts with $\varepsilon$, one (anti)meron moves outside the plot region in (a,e). Massless case would inherit the in-plane texture only and lead to a vortex-vortex pair. Black circle indicates the on-shell ring as in Fig.~\ref{Fig:skyrmion}. 
(Bottom) Intensity contour profile of the spin channel signal $|\brho(\omega,\bk)|$ at zero temperature is in one-to-one correspondence with the top panel and the white dashed circle indicates the on-shell ring.
Parameters are $\chi=1,m=0.13,\gamma_0=0.1,\gamma_1=0.06,\mu=0.5$.
}\label{Fig:meron}
\end{figure*}

Now we inspect the other representative case with $\bgamma=\gamma_1\hat{x}$.
In terms of the symmetry profile, the spin texture should no longer remain the circular symmetry in the same sense that two EPs select a direction.
In fact, we have from Eq.~\eqref{eq:p_nu_maintext}
\begin{equation}\label{eq:p_nu_gamma1_maintext}
\begin{split}
    \bp_{12}&=2(\gamma_0\varepsilon-\chi\gamma_1k_y)\bd_{12} + [\varepsilon^2-(d^2+\gamma_0^2-\gamma_1^2)] \bgamma_1\\
    p_{3}&=2(\gamma_0\varepsilon-\chi\gamma_1k_y)m,
\end{split}
\end{equation}
which consists of a \textit{meron-antimeron pair} \cite{Goebel2021} in $\bk$-space when $m\neq0$ as shown in Fig.~\ref{Fig:meron}. A meron is defined as a half skyrmion in the sense that its spins traverse through either the northern or the southern hemisphere. 
In the situation of relaxation due to magnetic impurity scattering, $\gamma_1$ often accompanies the doping of in-plane magnetization $m_x\hat{x}$, 
which, putting $m_x\sigma_1$ to Eq.~\eqref{eq:H0}, merely shifts the whole texture along $k_y$ by $\chi m_x$  (see SM \ref{Sec:theta}). 
With this understood, it is instructive and without loss of generality to keep using model Eq.~\eqref{eq:H0}, where the mass
can also be generated from magnetic proximity effect or using intrinsic magnetic topological insulators \cite{Tokura2019,Wang2021}. 
In fact, with in-plane $\gamma_1$ present, even the massless case has already picked up the most distinguishing feature as a spin texture of in-plane \textit{vortex-vortex pair}, directly inherited from the meron pair.

Hence, it suffices to focus on the meron case for clarity and completeness.
Of importance is %$\bK^\pm=\left(0,\frac{1}{\gamma_1}\left[\chi\gamma_0\varepsilon\pm\sqrt{(\gamma_0^2-\gamma_1^2)(\varepsilon^2+\gamma_1^2)+\gamma_1^2m^2}\right]\right)$ 
\begin{equation}
\begin{split}
    \bK^\pm&=\left(0,\frac{1}{\gamma_1}\left[\chi\gamma_0\varepsilon\pm\sqrt{(\gamma_0^2-\gamma_1^2)(\varepsilon^2+\gamma_1^2)+\gamma_1^2m^2}\right]\right)
\end{split}
\end{equation}
where the in-plane spin vanishes.
The meron (antimeron) centered at $\bK^{\mp\sgn(m)}$ has respectively
\begin{equation}\label{eq:N_sk_gamma1}
    N_\mathrm{sk}=\pm 1/2,\mathtt{p}=\pm \chi/2,\mathtt{v}=\chi,\mathtt{h}=\mp \pi/2
\end{equation}
according to the characterization of skyrmion. Here, as yet another example besides the $\gamma_3$ case, the emergence of meron or vortex pair again owes entirely to the NH relaxation effect; and it more drastically alters the conventional spin-momentum locking. 
The dashed borderline in between the pair in Fig.~\ref{Fig:meron} is $k_y=K_y^0=\frac{\gamma_0}{\gamma_1}\chi\varepsilon$, where the spins are aligned along $\hat{x}$-direction. Note that the meron pair is \textit{perpendicular} to the EP pair along $k_x$-axis, if exists at all, in the foregoing $H_\mathrm{eff}$. 
The physical mechanism of the NH topological structures in the present $\gamma_1$ case share some key properties of the foregoing $\gamma_3$ case. On one hand, the formation of a meron or vortex again makes use of the winding feature of original spin-momentum locking in its vorticity. On the other hand, the NH $\gamma_1$ relaxation enables the creation of two such solitons to form a pair in the momentum space, in analogy to the monopole pair creation along energy axis with $\gamma_3$. Now at each generic energy plane, one meron/vortex in the soliton pair is brought in from momentum infinity rather than the energy infinity with $\gamma_3$ while the other is created near the momentum origin (SM \ref{SecSM:spin_channel}).
We will further summarize and clarify the physical picture for general NH relaxations in Sec.~\ref{Sec:summary}.

We can take a further look at the energy or momentum dependence of the $\gamma_1$ case featured in SARPES detection. Finite energy $\varepsilon$-plane leads to a shift along $k_y$, easily seen from the borderline $K_y^0$, and asymmetric distortion of the meron pair in Fig.~\ref{Fig:meron}. The direction of such shift and distortion fully depends on $\sgn[\varepsilon]$, i.e., whether the photoemission signal comes from the region above or below the band midpoint. Figure~\ref{Fig:meron}(c) of $\varepsilon=0$ simply leads to a symmetry with respect to $k_x$-axis, which is the pair borderline in this energy plane.
Independent of energy planes, the large-momentum asymptotic form when $k\gg\varepsilon,\gamma_1, m$ reads %$\bp_\mathrm{as}=\gamma_1(k_y^2-k_x^2,-2\chi k_xk_y,0)$, 
\begin{equation}\label{eq:p_as}
    \bp_\mathrm{as}=\gamma_1(k_y^2-k_x^2,-2\chi k_xk_y,0),
\end{equation}
which is an in-plane $d$-wave-like nontrivial vortex texture of winding number $2\chi$. This is another observable consequence of the presence of two in-plane vortices of the same vorticity in Eq.~\eqref{eq:N_sk_gamma1}: at large momentum, the summed winding topology of the two persists while the inner structure at small momentum is concealed.

The on-shell large-$(\varepsilon,\bk)$ region produces $\bp_\mathrm{on}=2(\gamma_0\varepsilon-\chi\gamma_1k_y)\bd$, 
% \begin{equation}\label{eq:p_aspt2}
% \begin{split}
%     \bp=2(\gamma_0\varepsilon-\chi\gamma_1k_y)\bd
% \end{split}
% \end{equation}
proportional to the original spin-momentum locked texture as expected. However, compared to the previous $\gamma_0$ or $\gamma_3$ case, the $k_y$-dependence here interestingly receives a visible strength anisotropy within a certain energy plane. There remains the question how the topology of meron or vortex pair is connected to $\bp_\mathrm{on}$, which is roughly a single vortex. This is exactly explained by the foregoing $\varepsilon$-dependent shift and distortion of the profile. Starting from a particular energy plane displaying the meron pair, say, Fig.~\ref{Fig:meron}(c), we can then move the plane downward in order to reach an on-shell region for some large but finite $k$. During this procedure exemplified by Figs.~\ref{Fig:meron}(c,b,a), the pair borderline moves towards the negative $k_y$-direction and eventually one meron in the $k_y<0$ half-plane is pushed outside the on-shell observation window within $k$. The other one left as in Fig.~\ref{Fig:meron}(a), with associated distortion, gives rise to $\bp_\mathrm{on}$ around the on-shell ring region.

In the same manner as the previous $\gamma_3$ discussion, the bottom panel of Fig.~\ref{Fig:meron} shows the intensity profile. It breaks the symmetry between $k_x$ and $k_y$, which also happens in the density channel (see SM  Fig.~\ref{Fig:density_gamma1}). The $\varepsilon$-dependent movement of the strongest signal along the $k_y$-axis, from Fig.~\ref{Fig:meron}(a) to Fig.~\ref{Fig:meron}(e), is as well consistent with the movement of the meron or vortex pair described above. This makes one of the merons or vortices stronger in intensity but does not fundamentally affect the topological features. Note also that the region of a stronger signal is still consistent with part of the on-shell ring enhancement near the Dirac cone in Fig.~\ref{Fig:meron}(a,b,d,e), which is physically expected. Fig.~\ref{Fig:meron}(c) inside the gap is mainly contributed by the relaxation broadening and hence relatively weaker.

Together with the previous situation with $\gamma_3$, the results can have two important implications. i) The exceptional degeneracies and the associated bulk Fermi arc are not the directly and robustly observable objects in solid-state electronic photoemission, which is discussed in SM \ref{SecSM:density_channel} and above for respectively the density and spin channels. This is in stark contrast to most other NH platforms such as optics and other typically bosonic systems, where exceptional degeneracy can play a much more distinguishable role, and clearly reflects the uniqueness of solid-state NH electrons as we have emphasized.
In those other systems, retarded NH Hamiltonians are directly related to the diverging response due to exceptional degeneracy, while in electronic systems the convolution of retarded and advanced NH Green’s functions determines the physical responses.
ii) It is physically strongly required to go beyond the conventional exceptional degeneracy and hence the present distinct treatment provides the necessary method and viewpoint. New effects in the form of topological spin textures turn out to be the most relevant consequence of those NH relaxations; and, closely related to i), they are not qualitatively affected by the possible exceptional degeneracies and actually exist more stably in much wider parameter ranges.

\subsection{Critical direction of NH relaxation and topological vortex transition}

\begin{figure*}[!htbp]
\includegraphics[width=12.4cm]{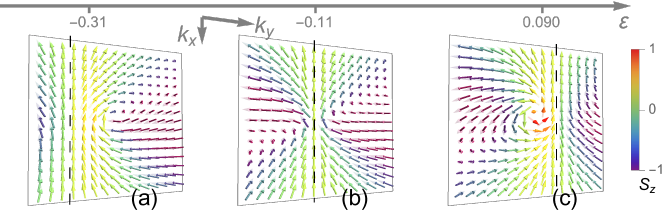}
\caption{Topological spin textures in the $(\varepsilon,\bk)$-space with $\bgamma$-relaxation in the direction $\theta=\pi/4$ measured from $\hat{z}$ towards $\hat{x}$. %Spins are in the globally rotated frame to facilitate presentation. 
(a,b,c) Three energy planes in the square-shaped $\bk$-space $[-1.1,1.1]^2$. One single vortex remains in each energy plane as the other vortex in Fig.~\ref{Fig:meron} or Fig.~\ref{Fig:theta>45} disappears at the infinity of $k_y$-axis. (b) is the special energy plane $\varepsilon^0$. Above $\varepsilon^0$ in (c), the spin-unidirectional dashed black line is on the right of the vortex along $k_y$-axis; the opposite happens in (a) below $\varepsilon^0$. %Other settings and parameters same as Fig.~\ref{Fig:theta>45}.
Parameters are $\chi=1,m=0.13,\gamma_0=0.1,\gamma_1=0.06$.
}\label{Fig:theta=45}
\end{figure*}

In the previous discussions, we focus on two physically representative and accessible situations: the NH relaxation vector is normal or in-plane to the surface. It is as well intriguing to understand what happens with intermediate cases, i.e., those with a general direction of $\bgamma=\gamma\sin{\theta}\hat{x}+\gamma\cos{\theta}\hat{z}$ with $\theta\in[0,\pi/2]$. 
With the typical relaxation mechanism of magnetic impurity scattering, this directly corresponds to a generic surface magnetization due to magnetic doping. Also, other directions can be readily related to this parametrization via symmetries and trivial rotation. 

Such a study helps identify the most stable features among what we highlighted above. 
Detailed in SM \ref{Sec:theta}, we find that the formation of monopole pair and vortex pair mostly persist and can coexist. 
Besides, the perfect skyrmion or meron does not exactly hold for general $\theta$ except in the previous two cases, especially because of the variation in the out-of-plane spin. Varying $\theta$ can actually modify the boundary condition at large momenta. For instance, the skyrmion number will deviate from an integer as we increase from $\theta=0$. However, distinguishing topological properties remain for general $\theta$, such as the quantized jump of (fractional) skyrmion number across the monopole energy planes and the topological winding numbers for the vortex pair. This is physically understandable because, as aforementioned, the monopole pair and vortex pair are much more robust objects and also form the core structures in the previous cases: the monopole generates the skyrmion charge while the vortex determines the in-plane texture of the meron.
More significantly, distinctively new features are hidden in Eq.~\eqref{eq:p_nu_maintext} with variable $\theta$, especially represented by a \textit{critical angle} $\theta=\pi/4$ and the topological transition from vortex-vortex pair to vortex-antivortex pair and also that of pair annihilation.

Therefore, the present NH system exhibits a significant and rare property: the distinct topological soliton objects can actually be deformed to each other, especially mediated by the topological transitions driven by tuning the direction of relaxation. 
Experimentally, this predicts an attractive dissipation-engineered topological phenomenon to be explored. 
Also, such a situation with stable and rich topological features becomes very helpful as it enables measurement and verification in more general settings and considerably relaxes the requirement in realistic detection, thus further promoting the potency and experimental relevance of the present proposal.

We first focus on the vortex aspect. %and use the notation $\tilde k_y=\chi k_y$ and $\tilde\bk=(k_x,\tilde k_y)$ for brevity. 
Within the wide range $\pi/4<\theta\leq\pi/2$, one observes the formation of a \textit{vortex-vortex pair} %centered at $\tilde\bK^s=(0,\tilde k_y^s),s=\pm1$
labelled by $s=\pm1$ and characterized by vorticity and helicity
\begin{equation}\label{eq:vt_character_>45_main}
    \mathtt{v}=\chi,\mathtt{h}=-s\frac{\pi}{2},
\end{equation}
which is the main feature and stably persists. It not only smoothly connects to the pure $\gamma_1$ ($\theta=\pi/2$) case with the vortex pair embedded in the meron pair, but also highly resembles its main features (see SM Fig.~\ref{Fig:theta>45} in close analogy to Fig.~\ref{Fig:meron}). For instance, there similarly exists a spin-unidirectional line %$\tilde K_y^0=\frac{\gamma_0 \varepsilon+\cos\theta\,\gamma d_3}{\gamma\sin\theta}$ 
$K_y^0=\chi\frac{\gamma_0 \varepsilon+\cos\theta\,\gamma d_3}{\gamma\sin\theta}$ 
as the borderline in between the vortex-vortex pair, although no longer exactly midway between the vortices. The counterpart of the $2\chi$-winding large-momentum asymptotic vortex texture Eq.~\eqref{eq:p_as} also exists.
To facilitates further description, we denote $\alpha(\varepsilon)=\gamma d_3+\cos\theta\,\gamma_0\varepsilon$ and a special energy plane $\varepsilon^0$ such that $\alpha(\varepsilon^0)=0$.
When we reduce $\theta$ to approach $\pi/4$, one vortex in the above vortex-vortex pair moves to infinity in momentum space while the other one remains. 
It is hence left with a \textit{single vortex} characterized by 
\begin{equation}\label{eq:vt_character_=45_main}
    \mathtt{v}=\chi,\mathtt{h}=\sgn(\alpha)\frac{\pi}{2},
\end{equation}
which is shown in Fig.~\ref{Fig:theta=45}.
Approaching the special $\varepsilon^0$ energy plane, the vortex moves to infinity in momentum space, explaining opposite-helicity vortices across the special energy plane with no vortex.

\begin{figure*}[!htbp]
\includegraphics[width=17.8cm]{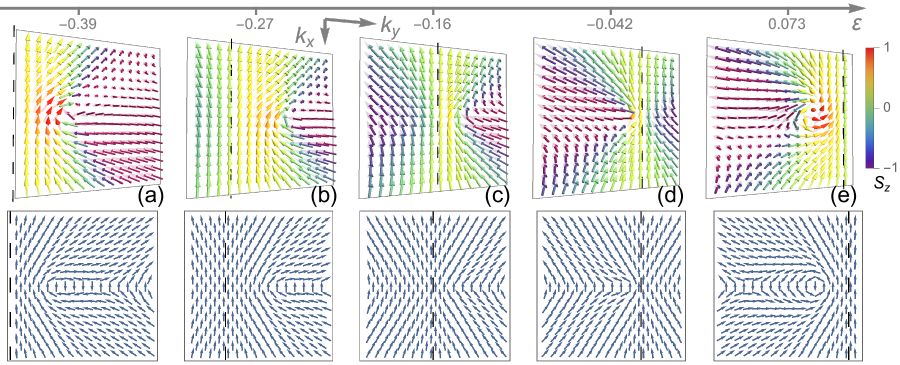}
\caption{Topological spin textures in the $(\varepsilon,\bk)$-space with $\bgamma$-relaxation in the direction $\theta=\pi/6$ measured from $\hat{z}$ towards $\hat{x}$.
Top: full spin vectors; Bottom: in-plane spin components only to guide eyes.
%Spins are in the globally rotated frame to facilitate presentation. 
Five energy planes in the square-shaped $\bk$-space (a) $[-1.0,2.1]^2$ and (b,c,d,e) $[-1.1,0.6]^2$. A vortex-antivortex pair appears in energy planes (a,e), where the vortex is inherited from Fig.~\ref{Fig:theta=45} and the antivortex comes in from the infinity of $k_y$-axis. This pair annihilates at critical energy planes (b,d) respectively at $\varepsilon=\varepsilon_\mathrm{vt}^\mp$, resulting in a trivial texture in (c). Dashed black line where spins are uniformly aligned no longer indicates the borderline between the pair since it lies to either side of the whole vortex-antivortex pair, if present. Other settings and parameters same as Fig.~\ref{Fig:theta=45}.}\label{Fig:theta<45}
\end{figure*}

This topological transition at the critical angle $\theta=\pi/4$ qualitatively changes the vortex profile and can be intuitively understood as the vortex-vortex pair cannot directly annihilate, in contrast to the situation below. 
In fact, reducing from $\theta=\pi/4$ with this single vortex, its antivortex partner can be created and come in also from the infinity. When $\theta<\pi/4$, there is a pair of critical energy planes $\varepsilon_\mathrm{vt}^\pm$, outside which a \textit{vortex-antivortex pair} labelled by $s=\pm1$ always exists and is characterized by
\begin{equation}\label{eq:vt_character_<45_main}
    \mathtt{v}=s\chi\sgn(\alpha),\mathtt{h}=\sgn(\alpha)\frac{\pi}{2}.
\end{equation}
As shown in Fig.~\ref{Fig:theta<45}, the whole pair lies along the $k_y$-axis either below %($\varepsilon>\varepsilon_\mathrm{vt}^+$) 
or above %($\varepsilon<\varepsilon_\mathrm{vt}^-$) 
the spin-unidirectional line, which becomes no longer a borderline inside a pair as in the $\theta>\pi/4$ case, also indicating the qualitative change across the critical $\theta=\pi/4$. Moving to the inner region between the two critical $\varepsilon_\mathrm{vt}^\pm$ planes, the pair annihilates at such planes as a distinct topological transition. This pair annihilation transition in the $(\varepsilon,\bk)$-space at any given $\theta<\pi/4$ is not to be confused with the $\theta=\pi/4$ topological transition of the overall texture profile.

On the other hand, a monopole-antimonopole pair at energy planes $\varepsilon_\mathrm{mp}^s$ with topological charge $C=\chi s$ for $s=\pm1$ always exists for $\theta<\pi/2$. 
Approaching the pure $\gamma_1$ case ($\theta=\pi/2$), the monopoles will move to the positive or negative energy plane at infinity, i.e., they will eventually merge into the bulk bands and disappear. The above topological transition at $\theta=\pi/4$ also entails corresponding boundary condition change at large momenta, naturally mediated by the foregoing creation and annihilation of (anti)vortices at infinity. The above $2\chi$-winding for $\theta>\pi/4$ vortex-vortex pair is thus changed to one without large-momentum winding for $\theta<\pi/4$ vortex-antivortex pair.
Given this trivial boundary configuration and further approaching the pure $\gamma_3$ case ($\theta=0$), the spins actually align uniformly in the out-of-plane direction at large momenta; hence a perfect skyrmion can be generated from a monopole as we discussed earlier.
Besides, the above vortex-antivortex pair becomes a redundant description because it is simply a mathematical artifact when $\theta=0$.
Other than these special cases, i.e., when $0<\theta<\pi/2$, coexistence of monopoles and vortices holds in general. 
It is interesting to compare the critical energy planes of vortex pair annihilation with the monopole energy planes when $0\leq\theta\leq\pi/4$, leading to an insightful relation
\begin{equation}\label{eq:epsilon_mp_vt_main}
    \varepsilon_\mathrm{mp}^-\leq\varepsilon_\mathrm{vt}^- \leq \varepsilon^0 \leq \varepsilon_\mathrm{vt}^+\leq\varepsilon_\mathrm{mp}^+.
\end{equation}
The inner two equalities hold when $\theta=\pi/4$; the outer two hold when $\theta=0$. This relation makes the $\theta$-dependent evolution transparent: the vortex annihilation planes $\varepsilon_\mathrm{vt}^\pm$ are created from the special $\varepsilon^0$-plane at the critical $\theta=\pi/4$ situation; they move apart and eventually exactly deform to the monopole pairs at $\varepsilon_\mathrm{mp}^\pm$ planes when $\theta$ is reduced and approaches 0; this makes the energy region in between the two planes featureless since vortex pair annihilation already occurs, consistent with our $\gamma_3$ description that skyrmions are outside such planes.

\section{Discussion and conclusion}\label{Sec:summary}

We summarize and discuss below the physical mechanism of the formation of new NH topological textures. One crucial aspect is that the relevant NH relaxation is spin-dependent and thus specifies a direction in spin space; spin-balanced NH relaxation alone like $\gamma_0$ cannot make this happen. This is in close connection to the surface state with SOI and magnetic impurities we adopted as the realistic platform. 
Such NH relaxations have a fundamental impact on the overall landscape of the spins in the energy-momentum space.
As previously described in $\gamma_3$ and $\gamma_1$ cases, the new NH topological solitons can make use of the in-plane winding feature from the original spin-momentum correlation in the surface state Dirac bands. 
Its combination with NH relaxations with a specified spin direction is essential and leads to the new solitons.
Intuitively, $\gamma_3$ specifies an out-of-plane direction and can promote the in-plane winding to circularly symmetric skyrmion and monopole textures with nontrivial out-of-plane variation; $\gamma_1$ selects an in-plane direction and exactly uses it as the borderline between the two merons/vortices placed perpendicular to that direction.

Moreover, the combination of the Dirac model and NH relaxations exhibits yet another important effect: the new solitons can be created and brought inside from the energy or momentum infinity, leading to the intriguing pair formation of monopole, meron and vortex in the NH scenario. Such a pair creation mechanism taking advantage of the energy-momentum infinity is essential to the whole phenomenon; it also plays an important role together with other topological transitions, where solitons can either move to disappear at infinity or annihilate, for the foregoing general direction of relaxation. 
In reality with higher-energy bulk bands, solitons at infinity are actually formed around the edge between surface and bulk bands and quickly approach the NH Dirac model prediction as they move inside the main surface state region.

We lastly discuss the detection in realistic SARPES measurements, for which we introduce the characteristic energy $\varepsilon'$, momentum $k'$, and relaxation strength $\gamma'$ and physically assume $\gamma_0\pm\gamma_i$ both of the order of magnitude of $\gamma'$. The energy $\varepsilon'$, typically given by the exchange gap $\sim40\textrm{-}80\mathrm{meV}$, can also be the finite energy $\varepsilon$ away from the band midpoint \cite{Tokura2019,Wang2021}. 
The spin relaxation time can be experimentally estimated to be  $3\textrm{-}12\mathrm{ps}$ from topological insulator surface state pumping measurements, which leads to $\gamma'\sim 0.4\textrm{-}2\mathrm{meV}$ and even larger with magnetic doping that enhances the magnetic impurity scattering \cite{Cacho2015,Iyer2018}.
Experimentally, the quantum anomalous Hall effect and other new phases related to magnetism are considered to be strongly affected by such magnetic disorder and the induced relaxation $\gamma'$, which, among other effects, gives the considerable density of states within the gap \cite{Tokura2019,Wang2021}. For instance, tunable and intense disorder at the nanoscale and hence large enough $\gamma'$ are generally confirmed for chromium-doped and other codoped topological insulators \cite{Lee2015,Liu2020,Tokura2019,Wang2021}. Such a situation is therefore very suitable for the experimental realization of the NH topological spin textures.
With $v\sim3\times10^5\mathrm{m/s}$, one can use $k'=\varepsilon'/v\sim 0.02\textrm{-}0.04\mathrm{\text{\AA}^{-1}}$ to represent an estimation of either the typical skyrmion size (from the center to the radius where spins point in-plane) or the meron pair separation measured between two centers (SM \ref{SecSM:estimation}).
A beneficial feature is that $\gamma'$ affects the signal strength but not directly the fine texture. The typical $\bk$-space scales of those fine textures will not become too small even when the relaxation $\gamma'$ are not large, which as well partially reflects the topological robustness of those textures. 
The probe pulse from synchrotron and especially laser light source with high photon flux and long duration can provide ultrahigh energy and momentum resolution up to $\lesssim1\mathrm{meV}$ and $0.005\textrm{-}0.01\mathrm{\text{\AA}^{-1}}$ that are well capable of observing the present phenomena \cite{Jozwiak2013,Cacho2015,Jozwiak2016,Reimann2018,Lv2019,Sobota2021}. Note that the capability of SARPES detection has already been substantially demonstrated, especially in equilibrium situations. 

In summary, we study the NH system of topological surface state with spin-dependent relaxation and relate it to concrete photoemission spectroscopy. Beyond the effective NH Hamiltonian approach, interesting new features manifest in the spin channel as various exotic topological soliton textures. They constitute the robust NH spectroscopic feature and fully surpass the existence of exceptional degeneracies. Such observable effects provide the much-needed experimental relevance for solid-state NH systems and renew the exploration of NH physics and topological states in contemporary spectroscopy.

\begin{acknowledgments}
We are thankful for the helpful discussion
with Y. Michishita and I. Belopolski. 
This work was supported by JSPS KAKENHI (No.~18H03676) and JST CREST (No. JPMJCR1874). X.-X.Z was also supported by RIKEN Special Postdoctoral Researcher program.
\end{acknowledgments}\mycomment{\Yinyang}

%\subsection*{Author contributions}
%Xiao-Xiao Zhang and Naoto Nagaosa conceptualized and designed the study. Xiao-Xiao Zhang carried out the calculation and analysis. All authors contributed to the discussion and writing of the manuscript.

%\nocite{*}

\bibliography{reference.bib}  % The references (bibliography) information are stored in the file named "Bibliography.bib"
\let\addcontentsline\oldaddcontentsline% Restore \addcontentsline

\newpage
\onecolumngrid
\newpage
{
	\center \bf \large 
	Supplemental Information \\
	%\large for ``Non-Hermitian exceptional Landau quantization in electric circuits"\vspace*{0.1cm}\\ 
	\large for ``\newtitle"\vspace*{0.1cm}\\ 
	\vspace*{0.5cm}
	%\newauthor
}
\begin{center}
    %\getauthor \\
	Xiao-Xiao Zhang$^1$ and Naoto Nagaosa$^{2,1}$\\
	\vspace*{0.15cm}
	\small{$^1$\textit{RIKEN Center for Emergent Matter Science (CEMS), Wako, Saitama 351-0198, Japan}}\\
	\small{$^2$\textit{Department of Applied Physics, University of Tokyo, Tokyo 113-8656, Japan}}\\
	\vspace*{0.25cm}	
\end{center}

%\twocolumngrid	

\tableofcontents

% %\clearpage
% %\appendix
% \setcounter{equation}{0}
% \setcounter{figure}{0}
% \setcounter{table}{0}
% \setcounter{page}{1}
% %\renewcommand{\theequation}{S\arabic{equation}}
% \renewcommand{\thefigure}{S\arabic{figure}}
% \renewcommand{\bibnumfmt}[1]{[S#1]}
% %\renewcommand{\citenumfont}[1]{S#1}

%%%%%%%%%% Merge with supplemental materials %%%%%%%%%%
%%%%%%%%% Prefix a "S" to all equations, figures, tables and reset the counter %%%%%%%%%%
%\appendix
\setcounter{section}{0}
\setcounter{equation}{0}
\setcounter{figure}{0}
\setcounter{table}{0}
\setcounter{page}{1}
%\makeatletter
\renewcommand{\theequation}{S\arabic{equation}}
\renewcommand{\thefigure}{S\arabic{figure}}
\renewcommand{\theHtable}{Supplement.\thetable}
\renewcommand{\theHfigure}{Supplement.\thefigure}
\renewcommand{\bibnumfmt}[1]{[S#1]}
\renewcommand{\citenumfont}[1]{S#1}
%%%%%%%%% Prefix a "S" to all equations, figures, tables and reset the counter %%%%%%%%%%

\section{Keldysh formalism and spectroscopy}\label{SecSM:Keldysh}
To calculate for the photoemission spectroscopy, we work along the Keldysh time contour with a forward '$+$' branch and a backward '$-$' branch. The nonequilibrium Green's function matrix reads
\begin{equation}
\hat{\cG}=\begin{bmatrix}
\cG^{++} & \cG^{+-} \\
\cG^{-+} & \cG^{--} 
\end{bmatrix}
=\begin{bmatrix}
\cG^{\mathds{T}} & \cG^< \\
\cG^> & \cG^{\tilde{\mathds{T}}} 
\end{bmatrix}
\end{equation}
and the Keldysh rotated one is
\begin{equation}\label{eq:KeldyshRotation}
    \check{\cG}=R\tau_3\hat{\cG}R^\dag=\begin{bmatrix}
\cG^\mathrm{r} & \cG^\mathrm{k} \\
0 & \cG^\mathrm{a}
\end{bmatrix}
\end{equation}
with $R=\frac{1}{\sqrt{2}}\begin{bmatrix}
1 & -1 \\
1 & 1
\end{bmatrix}$ and pauli matrix $\tau_3$ in the Keldysh space. The Dyson equation $\cG = \cG_0(1+ \Sigma \cG)$ can in general be derived in both cases.  Keldysh-space matrix multiplication and argument convolution is always implicitly assumed.
The exact Dyson equation of $\cG^<$ takes the form 
\begin{equation}\label{eq:G^<identity0}
    \cG^< = (1+\cG^\mathrm{r}\Sigma^\mathrm{r}) \cG_0^< (1+\Sigma^\mathrm{a}\cG^\mathrm{a}) + \cG^\mathrm{r} \Sigma^< \cG^\mathrm{a},
\end{equation}
where only the second part is nonvanishing for steady states \cite{Keldysh1}. Therefore, we henceforth rely on 
\begin{equation}\label{eq:G^<identity}
    \cG^< = \cG^\mathrm{r} \Sigma^< \cG^\mathrm{a}.
\end{equation}

We have for the retarded and advanced Green's functions
\begin{equation}
    \cG^\mathrm{r(a)}=\frac{1}{\omega-H_0-\Sigma_B^\mathrm{r(a)}} = \left( \frac{1}{\tilde{\varepsilon}\sigma_0-\tilde{d}\cdot\bsigma} \right)^{(\dag)} = \left( \frac{\tilde{\varepsilon}\sigma_0+\tilde{d}\cdot\bsigma}{\tilde{\varepsilon}^2-\tilde{d}^2} \right)^{(\dag)}
\end{equation}
where we start to use the notation $\tilde{\varepsilon}=\varepsilon+\ii\varepsilon'$, $\tilde{\bd}=\bd+\ii\bd'$,  $\varepsilon=\omega-d_0,\varepsilon'=\gamma_0,\bd'=-\bgamma$. 
And we derive the following general formula, after lengthy algebraic manipulation,
\begin{equation}\label{eq:r<a}
\begin{split}
    & (\tilde{\varepsilon}\sigma_0-\tilde{d}\cdot\bsigma)^{-1} (\zeta_\nu\sigma^\nu ) (\tilde{\varepsilon}^*\sigma_0-\tilde{d}^*\cdot\bsigma)^{-1} = \frac{1}{|\tilde{\varepsilon}^2-\tilde{d}^2|^2} \zeta_\nu \cA^\nu(\tilde{\lambda}) 
\end{split}
\end{equation}
where the 4-vector $\cA^\nu(\tilde{\lambda})$ is given by 
\begin{equation}\label{eq:cA0}
\begin{split}
    \cA_0 &=a_0^+\sigma_0 + \ba^+\cdot\bsigma \\
    \bcA &= \ba^-\sigma_0 + a_0^-\bsigma +  \dvec{t}\cdot\bsigma - \bn\times\bsigma  
\end{split}
\end{equation}
with the notation
\begin{equation}\label{eq:cA0_coef}
\begin{split}
    a_0^\pm = |\tilde{\varepsilon}|^2 \pm |\tilde{d}|^2,\qquad &\ba^\pm=2(\varepsilon \bd + \varepsilon'\bd'\pm\bd\times\bd') \\
    \dvec{t} = 2(\bd\bd+\bd'\bd'),\qquad &\bn=2 (\varepsilon'\bd-\varepsilon\bd')
\end{split}
\end{equation}
where $\dvec{t}$ is a rank-2 symmetric tensor.
Henceforth, we introduce the shorthand 4-vector notation $\tilde{\lambda}_\mu=(\tilde{\varepsilon},\tilde{\bd})$ and the quantity $\zeta_\nu$ in Eq.~\eqref{eq:r<a} is completely generic. As the main use, Eq.~\eqref{eq:r<a} helps us readily obtain
\begin{equation}\label{}
\begin{split}
    \cG^<(\omega) & =\cG^\mathrm{r} \Sigma_B^< \cG^\mathrm{a} =  \frac{2\ii f(\omega)}{|\tilde{\varepsilon}^2-\tilde{d}^2|^2} \gamma_\nu \cA^\nu (\tilde{\lambda})
\end{split}
\end{equation}
and
\begin{equation}\label{eq:p_nu}
\begin{split}
    2p_\mu&=\mathrm{Tr}[\sigma_\mu\gamma_\nu \cA^\nu] \\
    p_0& = \gamma_0(\varepsilon^2+E^2) + 2 \varepsilon\bd\cdot\bgamma \\
    \bp &= 2(\gamma_0\varepsilon+\bd\cdot\bgamma)\bd + (\varepsilon^2-E^2) \bgamma
\end{split}
\end{equation}
with $E^2=d^2+\gamma_0^2-\gamma^2$, which is Eq.~\eqref{eq:p_nu_maintext} in the main text.

The SARPES signal matrix for the time-resolved case has been shown to have the expression \cite{Freericks2009,Zhang2022}
\begin{equation}
\begin{split}
    P(\omega,t) &= -\mycomment{\frac{\ii}{\hbar^2}}\ii\int\mycomment{_{-\infty}^{\infty}} \mycomment{\int\mycomment{_{-\infty}^{\infty}}} \dd t_1 \dd t_2\, \ee^{\ii\varepsilon (t_1-t_2)} s(t_1-t)s(t_2-t) \cG^<(t_1, t_2)
\end{split}
\end{equation}
with $s(t)=(2\pi t_\mathrm{p}^2)^{-\frac{1}{2}}\ee^{-t^2/2 t_\mathrm{p}^2}$ the isotropic probe pulse of width $t_\mathrm{p}$. For the present equilibrium state, we have instead $\cG^<(t_1, t_2)\rightarrow\cG^<(t_1-t_2)$, leading to the time-independent form
\begin{equation}
\begin{split}
    P(\omega) &= \frac{-\ii}{2\pi}\int\dd \omega'\; \cG^<(\omega')\ee^{-(\omega-\omega')^2t_\mathrm{p}^2}
\end{split}
\end{equation}
and hence the component decomposed to every channel
\begin{equation}
\begin{split}
    P_\mu(\omega) &= \mathrm{Tr}[\sigma_\mu P] \\
    &= \frac{2}{\pi}\int\dd \omega' \ee^{-(\omega-\omega')^2t_\mathrm{p}^2} \frac{f(\omega')}{|\tilde{\varepsilon}(\omega')^2-\tilde{d}^2|^2}\,p_\mu(\omega').
\end{split}
\end{equation}
When the probe pulse is wide enough $t_\mathrm{p}/\sqrt{\pi}\,\ee^{-(\omega-\omega')^2t_\mathrm{p}^2}\rightarrow \delta(\omega-\omega')$, the Gaussian blurring becomes trivial and we can simply focus on $\cG^<$
\begin{equation}
\begin{split}
    P_\mu(\omega) 
    &\approx \frac{2}{\sqrt{\pi}t_\mathrm{p}} \frac{f(\omega)}{|\tilde{\varepsilon}(\omega)^2-\tilde{d}^2|^2}\,p_\mu(\omega).
\end{split}
\end{equation}

\section{Generation of general NH dissipation}\label{SecSM:photoemission}

\subsection{General dissipation from bath environment}\label{SecSM:bath}
Now we show how a fermionic bath environment can induce the general form of self-energies given in the main text. We start from the interacting Hamiltonian
\begin{equation}
    H=h_0+h_B.
\end{equation}
The bare surface state electrons have, according to Eq.~\eqref{eq:H0},
\begin{equation}
    h_0 =  \sum_\bk \psi_\bk^\dag H_0(\bk) \psi_\bk = \sum_{\bk,a=\pm} \varepsilon_{\bk a} \psi_{\bk a}^\dag\psi_{\bk a}
\end{equation}
where $\psi_\bk^\dag=\psi_{\bk\alpha}^\dag=(\psi_{\bk\uparrow}^\dag,\psi_{\bk\downarrow}^\dag)$. The fermionic bath and the interaction takes the general form 
\begin{equation}\label{eq:h_B}
    h_B =  \sum_{i\alpha\bk} [c_i^\dag H_i c_i + ( \psi_{\bk}^\dag V_{i} c_i + \mathrm{H.c.} )]
\end{equation} 
where $c_i^\dag=c_{i\alpha}^\dag=(c_{i\uparrow}^\dag,c_{i\downarrow}^\dag)$. 
The bath fermions' free Keldysh Green's functions are
\begin{equation}\label{eq:g_B}
\begin{split}
    g_i^{\mathrm{r}(\mathrm{a})}(\omega) & = (\omega\pm\ii0^+-H_i)^{-1} \\
    g_i^<(\omega) &=  2\pi \ii f(\omega) \delta(\omega-H_i).
\end{split}
\end{equation} 
Without loss of generality, we specify the coupling matrix to be
\begin{equation}\label{eq:V_i}
    V_i=\begin{bmatrix}
        t_{i\uparrow} & \\
         & t_{i\downarrow}
    \end{bmatrix}.
\end{equation}

We first consider that $H_i$ is diagonal in the spin-$z$ basis
\begin{equation}
    H_i=\begin{bmatrix}
        \varepsilon_{i\uparrow} & \\
         & \varepsilon_{i\downarrow}
    \end{bmatrix},
\end{equation}
which could imply that the bath system is spin-polarized to the $z$ direction in its ground state.
Then we have from Eq.~\eqref{eq:g_B}
\begin{equation}\label{eq:g_i1}
\begin{split}
        g_i^{\mathrm{r}(\mathrm{a})}(\omega) & = \begin{bmatrix}
        \omega\pm\ii0^+-\varepsilon_{i\uparrow} & \\
         &  \omega\pm\ii0^+-\varepsilon_{i\downarrow}
    \end{bmatrix}^{-1} \\ 
        &= \begin{bmatrix}
        \cP(\omega-\varepsilon_{i\uparrow})^{-1}\mp\ii\pi\delta(\omega-\varepsilon_{i\uparrow}) & \\
         &  \cP(\omega-\varepsilon_{i\downarrow})^{-1}\mp\ii\pi\delta(\omega-\varepsilon_{i\downarrow})
    \end{bmatrix}
\end{split}
\end{equation}
where the Sokhotski–Plemelj relation is used 
and
\begin{equation}\label{}
\begin{split}
        g_i^<(\omega) & = 2\pi\ii f(\omega) \begin{bmatrix}
        \delta(\omega-\varepsilon_{i\uparrow}) & \\
         &  \delta(\omega-\varepsilon_{i\downarrow})
    \end{bmatrix} .
\end{split}
\end{equation}
One can define a set of constant linewidth function in the wide-band limit \cite{Keldysh1,Stefanucci2015}
\begin{equation}\label{eq:gamma_alpha}
    \gamma_\alpha = \pi\sum_i |t_{i\alpha}|^2\delta(\varepsilon-\varepsilon_{i\alpha}).
\end{equation}
Then, using Eq.~\eqref{eq:g_i1}, we find the self-energies from Eq.~\eqref{eq:V_i} and the coupling in Eq.~\eqref{eq:h_B}
\begin{equation}
\begin{split}
    \Sigma_B^{\mathrm{r}(\mathrm{a})}(\omega) & = \sum_i V_i g_{i}^\mathrm{r(a)} V_i^\dag =\mp\ii \begin{bmatrix}
        \gamma_{\uparrow} & \\
         &  \gamma_{\downarrow}
    \end{bmatrix} =\mp\ii(\gamma_0\sigma_0+\gamma_3\sigma_3) 
\end{split}
\end{equation}
where $\gamma_0=(\gamma_{\uparrow} + \gamma_{\downarrow})/2,\gamma_3=(\gamma_{\uparrow} - \gamma_{\downarrow})/2$. The summation with principal values is guaranteed to vanish under the wide-band limit and we have as well used Eq.~\eqref{eq:gamma_alpha}.
Similarly, we can obtain 
\begin{equation}
\begin{split}
        \Sigma_B^<(\omega) & = \sum_i V_i g_{i}^< V_i^\dag =  2\ii f(\omega)(\gamma_0\sigma_0+\gamma_3\sigma_3) .
\end{split}
\end{equation}
Note also that we always have $\gamma_0>\gamma_3$ since $\gamma_\alpha>0$, as stated in the main text.

Next, we consider that $H_i$ is diagonal in the spin-$x$ basis
\begin{equation}
    H_i=\begin{bmatrix}
        \varepsilon_i & \Delta_{i}  \\
          \Delta_{i} & \varepsilon_i
    \end{bmatrix} = S \begin{bmatrix}
        \varepsilon_{i+} & \\
         & \varepsilon_{i-}
    \end{bmatrix}S^{-1}
\end{equation}
where the unitary matrix $S$ diagonalizes $H_i$ with $\varepsilon_{i\pm}=\varepsilon_i\pm\Delta_i$. This could imply that the bath system is spin-polarized to the $x$ direction in its ground state. Also, in the present case, it is plausible to assume henceforth that $t_{i\uparrow}=t_{i\downarrow}=t_i$, because the spin states $\ket{\alpha=\uparrow,\downarrow}$ are on the same footing with respect to the bath spin states $\ket{a=\pm}$ polarized to the $x$ direction.
Then we have from Eq.~\eqref{eq:g_B}
\begin{equation}\label{eq:g_i2}
\begin{split}
        g_i^{\mathrm{r}(\mathrm{a})}(\omega) & = S \begin{bmatrix}
        \omega\pm\ii0^+-\varepsilon_{i+} & \\
         & \omega\pm\ii0^+-\varepsilon_{i-}
    \end{bmatrix}^{-1}S^{-1} \\ 
        &= \frac{1}{2}\left[(\omega\pm\ii0^+-\varepsilon_{i+})^{-1}+(\omega\pm\ii0^+-\varepsilon_{i-})^{-1}\right]\sigma_0 + \frac{1}{2}\left[(\omega\pm\ii0^+-\varepsilon_{i+})^{-1}-(\omega\pm\ii0^+-\varepsilon_{i-})^{-1}\right]\sigma_1.
\end{split}
\end{equation}
and
\begin{equation}\label{}
\begin{split}
        g_i^<(\omega) & =  2\pi\ii f(\omega) S \begin{bmatrix}
        \delta(\omega-\varepsilon_{i+}) & \\
         &  \delta(\omega-\varepsilon_{i-})
    \end{bmatrix} S^{-1} \\ 
        &= \pi\ii f(\omega)\left\{ \left[\delta(\omega-\varepsilon_{i+})+\delta(\omega-\varepsilon_{i-})\right]\sigma_0 + \left[\delta(\omega-\varepsilon_{i+})-\delta(\omega-\varepsilon_{i-})\right]\sigma_1 \right\}.
\end{split}
\end{equation}
One can again define a set of constant linewidth function in the wide-band limit for $a=\pm$
\begin{equation}\label{eq:gamma_alpha2}
    \gamma_a = \pi\sum_i |t_{i}|^2\delta(\varepsilon-\varepsilon_{ia}).
\end{equation}
Then, using Eq.~\eqref{eq:g_i2} and Eq.~\eqref{eq:gamma_alpha2}, in the similar manner as the above, we find the self-energies from Eq.~\eqref{eq:V_i} and the coupling in Eq.~\eqref{eq:h_B}
\begin{equation}
\begin{split}
    \Sigma_B^{\mathrm{r}(\mathrm{a})}(\omega) & = \sum_i V_i g_{i}^\mathrm{r(a)} V_i^\dag \\ 
    &= \mp\frac{\ii\pi}{2}\sum_i  \begin{bmatrix}
        |t_{i\uparrow}|^2[\delta(\omega-\varepsilon_{i+})+\delta(\omega-\varepsilon_{i-})] & t_{i\uparrow}t_{i\downarrow}^*[\delta(\omega-\varepsilon_{i+})-\delta(\omega-\varepsilon_{i-})]\\
        t_{i\uparrow}^*t_{i\downarrow}[\delta(\omega-\varepsilon_{i+})-\delta(\omega-\varepsilon_{i-})] &  |t_{i\downarrow}|^2[\delta(\omega-\varepsilon_{i+})+\delta(\omega-\varepsilon_{i-})]
    \end{bmatrix}  \\
    &= \mp\frac{\ii}{2}\begin{bmatrix}
        \gamma_+ + \gamma_- & \gamma_+ - \gamma_- \\
         \gamma_+ - \gamma_- &  \gamma_+ + \gamma_-
    \end{bmatrix} \\
    &=\mp\ii(\gamma_0\sigma_0+\gamma_1\sigma_1) 
\end{split}
\end{equation}
where $\gamma_0=(\gamma_{+} + \gamma_{-})/2,\gamma_1=(\gamma_{+} - \gamma_{-})/2$.
Similarly, we can also obtain 
\begin{equation}
\begin{split}
        \Sigma_B^<(\omega) & = \sum_i V_i g_{i}^< V_i^\dag =  2\ii f(\omega)(\gamma_0\sigma_0+\gamma_1\sigma_1) .
\end{split}
\end{equation}
Note also that we again always have $\gamma_0>\gamma_1$ since $\gamma_a>0$, as stated in the main text.

\subsection{Magnetic impurity scattering}\label{SecSM:imp}
We now exemplify the generation of NH relaxation in the presence of magnetic impurities, which is corresponding to the case of magnetic doping onto the topological insulator surface state. For simplicity, we consider the following short-ranged matrix-formed potential due to the impurities 
\begin{equation}
    U(\br) = \sum_i^N u_\nu\sigma^\nu\delta(\br-\bR_i),
\end{equation}
which assumes random uniform distribution $u_\nu$ and the disorder averaged correlation $\braket{U(\br)U(\br')}_\mathrm{imp}\sim\delta(\br-\br')$. The impurity potential strength is $u_\nu=(u_0,\bu)$ with $\bu$ typically for magnetic impurities. The impurity concentration is $n=N/A$ with $A$ the surface area of the system. We also assume that distinct types of impurity scattering is not correlated. Then we can study the disorder averaged retarded electronic Green's function as an example, which follows the Dyson equation
\begin{equation}\label{eq:g_imp}
    \braket{\cG^\mathrm{r}}_\mathrm{imp} = \cG_0^\mathrm{r} + \cG_0^\mathrm{r}\,\Sigma_\mathrm{i}^\mathrm{r} \braket{\cG^\mathrm{r}}_\mathrm{imp}
\end{equation}
where the bare surface state electronic Green's function is 
\begin{equation}
    \cG_0^\mathrm{r} = \frac{1}{\omega-H_0+\ii\eta} 
\end{equation}
with $H_0$ given by Eq.~\eqref{eq:H0} and infinitesimally small and positive $\eta$.

We neglect the quantum interference effect of impurity scattering due to cross diagrams. As the first nontrivial contribution to the relaxation effect, we can consider the first Born approximation for the self-energy $\Sigma_\mathrm{i}^\mathrm{r}$ due to impurity scattering, leading to the expression 
\begin{equation}
    \Sigma_\mathrm{i}^\mathrm{r}=\sum_\bk \frac{n}{A} u_{\nu_1} \sigma^{\nu_1}  \cG_0^\mathrm{r} u_{\nu_2} \sigma^{\nu_2}.
\end{equation}
Since the imaginary part of $\Sigma_\mathrm{i}^\mathrm{r}$ is relevant to giving rise to the NH relaxation, we look at 
\begin{equation}
\begin{split}
    \mathrm{Im} \cG_0^\mathrm{r} = \mathrm{Im} \frac{\varepsilon\sigma_0+\bd\cdot\bsigma}{\varepsilon^2-d^2-\eta^2+2\ii\varepsilon\eta}
    =-\frac{\pi}{2\varepsilon}\sum_{s=\pm1}\delta(\varepsilon-sd)\,(\varepsilon\sigma_0+\bd\cdot\bsigma)
\end{split}
\end{equation}
where the Sokhotski–Plemelj relation $1/(x\pm\ii\eta)=\cP\frac{1}{x}\mp\pi\ii\delta(x)$ and properties of Dirac delta function are used. This helps us to evaluate 
\begin{equation}\label{eq:sumImcG0}
\begin{split}
    \sum_\bk \mathrm{Im} \cG_0^\mathrm{r} = -\frac{\pi}{2} \xi_\mu\sigma^\mu.
\end{split}
\end{equation}
We notice that $\xi_{1,2}$ are zero since $d_{1,2}$ are linear in momentum. Then we find that $\xi_0=AD(\varepsilon)$ with the density of state $D$ and $\xi_3=m\xi_0/\varepsilon$ are the nonvanishing ones. Note also that with the more accurate self-consistent Born approximation beyond this simple calculation, the density of state $D$ and hence $\xi_0$ can gain a finite value even at the Dirac node for the massless case \cite{Bruus2004,Burkov2011}.
Because of the Pauli decomposition form in Eq.~\eqref{eq:sumImcG0}, it is beneficial to find the following formula
\begin{equation}\label{eq:cB0}
    u_{\nu_1} \sigma^{\nu_1}  \xi_\mu\sigma_\mu u_{\nu_2} \sigma^{\nu_2} = \xi_\mu \cB^\mu
\end{equation} 
with the 4-vector-like matrix $\cB^\mu$ given by \begin{equation}\label{eq:cB}
\begin{split}
    \cB_0 &=(u_0^2+u^2)\sigma_0 + 2u_0\bu\cdot\bsigma \\
    \bcB &= 2u_0\bu\sigma_0 + (u_0^2-u^2)\bsigma +  2\bu(\bu\cdot\bsigma).  
\end{split}
\end{equation}
To proceed, we consider the case with $\bu=u_3\hat{z}$, i.e., the magnetic impurity scattering is unbalanced between spin up and down, which is typically induced by impurity magnetization polarized in $\hat{z}$-direction. This example can be readily generalized to other situations by a rotation in the spin space. Now, only $\cB_0=(u_0^2+u_3^2)\sigma_0 + 2u_0u_3\sigma_3$ and $\cB_3=2u_0u_3\sigma_0+(u_0^2+u_3^2)\sigma_3$ are relevant.
From Eqs.~\eqref{eq:sumImcG0}\eqref{eq:cB0}, we find that 
\begin{equation}
    \mathrm{Im}\Sigma^\mathrm{r}_\mathrm{i}=-\frac{n\pi}{2A}\{[(u_0^2+u_3^2)\xi_0+2u_0u_3\xi_3]\sigma_0+[(u_0^2+u_3^2)\xi_3+2u_0u_3\xi_0]\sigma_3\}.
\end{equation}
Therefore, we have generated the $\sigma_0$ and $\sigma_3$ matrix structure of $\mathrm{Im}\Sigma^\mathrm{r}_\mathrm{i}=\gamma_0\sigma_0+\gamma_3\sigma_3$. Within the present approximation, we have in general $\xi_0>\xi_3$ and $\gamma_0>\gamma_3$ is guaranteed. Note that this stability condition holds well beyond this approximation, as is a general physical requirement for equilibrium systems. In the simpler case when $\xi_3=0$, e.g., for a spin degenerate band or the massless case, we directly have $\gamma_0 \propto u_0^2 + u_3^2$ and $\gamma \propto 2u_0 u_3$. Finally, with such a matrix structure of the relaxation, one can obtain from Eq.~\eqref{eq:g_imp} the dressed electronic Green's function as used in the main text.

\section{NH topological phases}\label{SecSM:NHtopo}
Here, we comment on the topological classification based on the effective NH Hamiltonian $H_\mathrm{eff}$ given in the main text. The essential methodology is to distinguish the type of gap in the complex energy plane in the first place. A gap at complex energy $E_0$ for a NH system $H_\mathrm{eff}$ means that its spectrum does not cross $E_0$, i.e.,  
\begin{equation}
    \Det(H_\mathrm{eff}(\bk)-E_0)\neq0 \quad \forall\bk\in \cM
\end{equation}
where $\cM$ is a certain set of points in the momentum space. Note that although $\cM$ can typically be the whole momentum space, it can also be a certain submanifold as per the context.
There are in general two types of gaps. 
\begin{itemize}
    \item Point gap: it is defined by the foregoing $E_0$. 
    \item Line gap: there exists a continuum of $E_0$, forming a line in the complex energy plane. Note that line gap implies point gap but not the converse.
\end{itemize}
Given the nature of these definitions, there is no simple-minded notion of being gapless. Instead, one always needs to specify the reference energy $E_0$ and the manifold $\cM$, with respect to which a NH system is gapless. For instance, the most prominent 'gapless' case for a multiband NH system manifests in the presence of any kind of exceptional degeneracies, i.e., setting $E_0$ to one such degeneracy $E_\mathrm{EP}$, where more than one bands touch and hence it is obviously gapless when $\cM$ is the whole momentum space.

Below, we discuss the NH topology for our particular model.
\begin{itemize}
    \item With exceptional degeneracy
    
This is the special 'gapless' case aforementioned. However, once we restrict $\cM$ to a subspace, say, an $S^p$ ($p=1$) loop around $E_\mathrm{EP}$, it is, in fact, only line-gap closed but still point-gapped, because the line crosses the loop while $E_\mathrm{EP}$ merely lies inside. Note that we obviously put the line of the line gap across the chosen degeneracy $E_\mathrm{EP}$. The context of being point-gapped is exactly how the topology is to be specified for those 'gapless' situations with exceptional degeneracy. 

\begin{itemize}
    \item a pair of EPs when $\gamma_1>|m|$ for the $\gamma_1$ case
    
The topological number is given by the spectral winding
\begin{equation}
    W_\mathrm{EP} = \frac{1}{2\pi}\oint_{S^1} \dd\bk \cdot\nabla_\bk \ln \Det[H_\mathrm{eff}(\bk)-E_0].
\end{equation}
This $W_\mathrm{EP}\in\mathbb{Z}$ evaluates to $\pm1$ in our case.

    \item an ER when $m=0$ for the $\gamma_3$ case
    
This massless case is protected by the chiral symmetry $C H_\mathrm{eff}^\dag(\bk)C^{-1}=-H_\mathrm{eff}(\bk)$ with $C=\sigma_3$. The winding loop $S^p$ ($p=0$) reduces to a pair of points inside and outside the ER, where the zeroth Chern number, i.e., number of occupied states, $n_\mathrm{i/o}$ is defined \cite{Okugawa2019,Yoshida2019}. This leads to the topological number $W_\mathrm{ER}=n_\mathrm{o}-n_\mathrm{i}\in\mathbb{Z}$ responsible for the ER. We have $W_\mathrm{ER}=1$ since inside (outside) the ER the spectrum is purely imaginary (real).
\end{itemize}

\item Fully line-gapped

On the other hand, the line gap is more akin to the Hermitian concept of a gap. For our $H_\mathrm{eff}$ in the main text, it possesses a line gap as long as no exceptional degeneracy exists, which is the \textit{most general} situation complementary to the more special EP/ER case. 
Indeed, the ER case requires exact masslessness while the EP case requires larger $\gamma_1$ than the mass.
In fact, with $\gamma_1$ present, we have the line gap along the real axis, which will be closed when $\gamma_1$ is reduced to $|m|$. And with $\gamma_3$ present, we have the line gap along either the real or imaginary axis, which will be closed when the system approaches the massless case.
In the absence of exceptional degeneracies, one can define the topology from the NH Berry curvature
\begin{equation}
    B_{n,ij}^{\alpha\beta}(\bk) = \ii \braket{\partial_i\psi_n^\alpha(\bk)|\partial_j\psi_n^\beta(\bk)},
\end{equation}
which leads to the NH Chern number 
\begin{equation}
    N_{n} = \frac{1}{2\pi}\int \epsilon_{ij}\dd^2\bk\; B_{n,ij}^{\alpha\beta}(\bk).
\end{equation}
Here, $n$ is the band index, $i=1,2$ means momentum components $k_i$, and $\alpha=L,R$ signifies the left/right eigenstate defined respectively from $H_\mathrm{eff}^\dag$ and $H_\mathrm{eff}$. It is important to note that the Chern number $N_n\in\mathbb{Z}$ does not depend on the choice of $\alpha,\beta$. For the lower band in our two-band model, the NH Chern number can be reduced to a simpler expression
\begin{equation}
    N = \frac{1}{4\pi}\int \frac{\dd^2\bk}{\tilde{d}^{3}} \tilde\bd\cdot(\partial_1\tilde\bd \times \partial_2\tilde\bd),
\end{equation}
which evaluates to $N=\sgn(\chi m)\frac{1}{2}$ for our model. The half-integral appearance is merely due to the unregularized continuum model we used.

\end{itemize}

\section{Analysis of the density and spin textures}\label{SecSM:Analyze_profiles}
% For the convenience of later discussion, we repeat the formula of the charge and spin channel signal given in the main text
% \begin{equation}\label{eq:p_nu}
% \begin{split}
%     2p_\mu&=\mathrm{Tr}[\sigma_\mu\gamma_\nu \cA^\nu] \\
%     p_0& = \gamma_0(\varepsilon^2+E^2) + 2 \varepsilon\bd\cdot\bgamma \\
%     \bp &= 2(\gamma_0\varepsilon+\bd\cdot\bgamma)\bd + (\varepsilon^2-E^2) \bgamma
% \end{split}
% \end{equation}
% with $E^2=d^2+\gamma_0^2-\gamma^2$.
\subsection{Zero-dissipation limit}\label{SM:nogamma}
\begin{figure*}[!htbp]
\includegraphics[width=10.4cm]{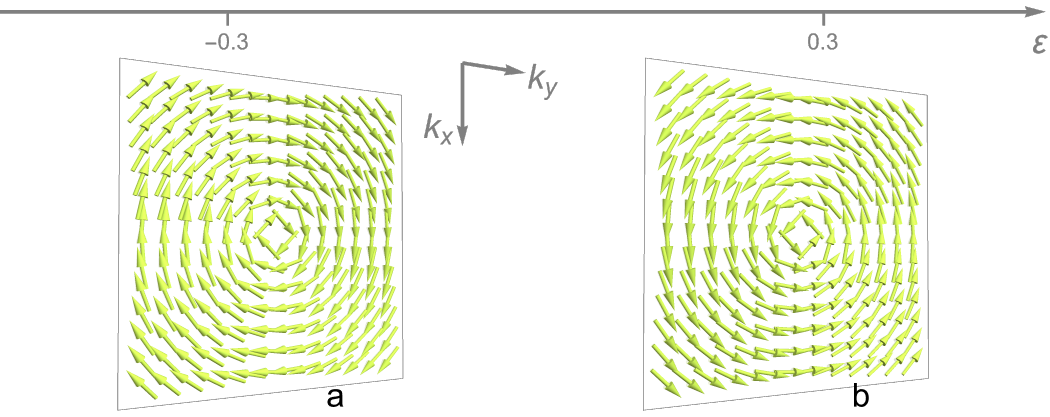}
\caption{Typical spin textures in the $(\varepsilon,\bk)$-space for the Hermitian topological insulator surface state.  
(a,b) Two energy planes in the square-shaped $\bk$-space $[-0.8,0.8]^2$. One single vortex-like winding structure represents the conventional spin-momentum locking. The winding sense is reversed across $\varepsilon=0$, where the signal vanishes and hence we do not show.
%Parameters are $\chi=1,m=0,\gamma_0=0.1,\gamma_1=0.06$.
}\label{Fig:hermitian}
\end{figure*}

To inspect the zero-dissipation limit, we discuss for simplicity the conventional case with $\gamma_0$ only and approach the limit of $\gamma_0\rightarrow0$.
Firstly, Eq.~\eqref{eq:p_nu} gives \begin{equation}\label{eq:p_gamma0}
\begin{split}
    p_0=\gamma_0(\varepsilon^2+E^2),\quad\bp= 2\gamma_0\varepsilon\bd
\end{split}
\end{equation}
where the original surface-state spin-momentum locking manifests in $\bp$. When $\gamma_0\rightarrow0$, Eq.~\eqref{eq:p_gamma0} apparently vanishes, which seems contradictory to the physical situation where the SARPES result should even exist with zero dissipation. This is resolved by noting that neither Eq.~\eqref{eq:p_nu} nor  Eq.~\eqref{eq:p_gamma0} is the expression in the first place, which is not suitable to discuss the limit. In fact, one can write down according to Eq.~\eqref{eq:G^<identity}
\begin{equation}
\begin{split}
    G^<=G^\mathrm{r}\Sigma^<G^\mathrm{a}&= \frac{\omega-h-\ii\gamma_0}{(\omega-h)^2+\gamma_0^2}\;\ii\gamma_0 f(\omega)\; \frac{\omega-h+\ii\gamma_0}{(\omega-h)^2+\gamma_0^2} 
\end{split}
\end{equation}
where $h$ is a generic Hamiltonian matrix representing, e.g., Eq.~\eqref{eq:H0}. Evidently, $G^\mathrm{r(a)}$ diverges as $G^\mathrm{r(a)}\sim\mp\ii\gamma_0^{-1}$ as long as $\Det[\omega-h]=0$. This immediately leads to 
\begin{equation}
\begin{split}
    \lim_{\gamma_0\rightarrow0} G^< = \ii f(\omega)\delta(\omega-h) 
\end{split}
\end{equation}
where the spectrum of $h$ manifests as $\delta$-function peaks. 
In Fig.~\ref{Fig:hermitian}, we directly plot the typical spin texture of the Hermitian case, i.e., the original topological insulator surface state, in comparison to the NH cases, e.g.,  Fig.~\ref{Fig:skyrmion} and Fig.~\ref{Fig:meron} in the main text. Obviously, one can observe one single winding structure representing the spin-momentum locking in the surface state. The winding sense is reversed across the zero-energy plane where the signal vanishes, which can be readily seen from the $\varepsilon$ factor in Eq.~\eqref{eq:p_gamma0}.
Besides, it does not bear any further nontrivial energy or momentum dependence, hence we only show two representative energy planes below or above the band midpoint. In contrast, we have the appearance of monopoles at specific energy planes in Fig.~\ref{Fig:skyrmion} and the meron/vortex-pair shift and deformation in the momentum space in Fig.~\ref{Fig:meron}.

\subsection{Density channel profile}\label{SecSM:density_channel}
Here, we discuss the spectroscopic signal in the less nontrivial density channel, where we rely on $\rho_0$ defined in the main text
\begin{equation}
    \rho_0  =  \frac{4 f(\omega)}{|\tilde{\varepsilon}^2-\tilde{d}^2|^2} p_0
\end{equation}
with $p_0$ given in Eq.~\eqref{eq:p_nu}. 
We have for the denominator
\begin{equation}
\begin{split}
    |\tilde{\varepsilon}^2-\tilde{d}^2|^2 & = |(\varepsilon+\ii\gamma_0)^2-(\bd-\ii\bgamma)^2|^2 \\
    &= (\varepsilon^2-E^2)^2 + 4(\varepsilon\gamma_0+\bd\cdot\bgamma)^2
\end{split}
\end{equation}
with $E^2=d^2+(\gamma_0^2-\gamma^2)$.
We first note that $\rho_0$ is \textit{positive-definite} as long as $\gamma_0>\gamma$ as mentioned in the main text. This is just what one would physically expect for a photoemission signal in the density/charge channel from a normal system in equilibrium. It suffices to notice that
\begin{equation}
\begin{split}
    p_0& = \gamma_0(\varepsilon^2+E^2) + 2 \varepsilon\bd\cdot\bgamma \\
    &= \gamma_0\left[\varepsilon^2+d^2 + 2 \varepsilon d\tilde\gamma\cos\theta +(\gamma_0^2-\gamma^2)\right]\\
    &\geq \gamma_0\left[\varepsilon^2+d^2 - |2 \varepsilon d\tilde\gamma| +(\gamma_0^2-\gamma^2)\right]\\
    &> \gamma_0\left[(\varepsilon-d)^2 +(\gamma_0^2-\gamma^2)\right]>0
\end{split}
\end{equation}
where we denote $\tilde\gamma=\gamma/\gamma_0<1$ and the angle $\theta$ between $\bd$ and $\bgamma$. Hence, the lack of a sharp feature such as sign change in the profile of the density channel strongly restricts any efficient experimental identification. 

\begin{figure}[hbt]
\includegraphics[width=16cm]{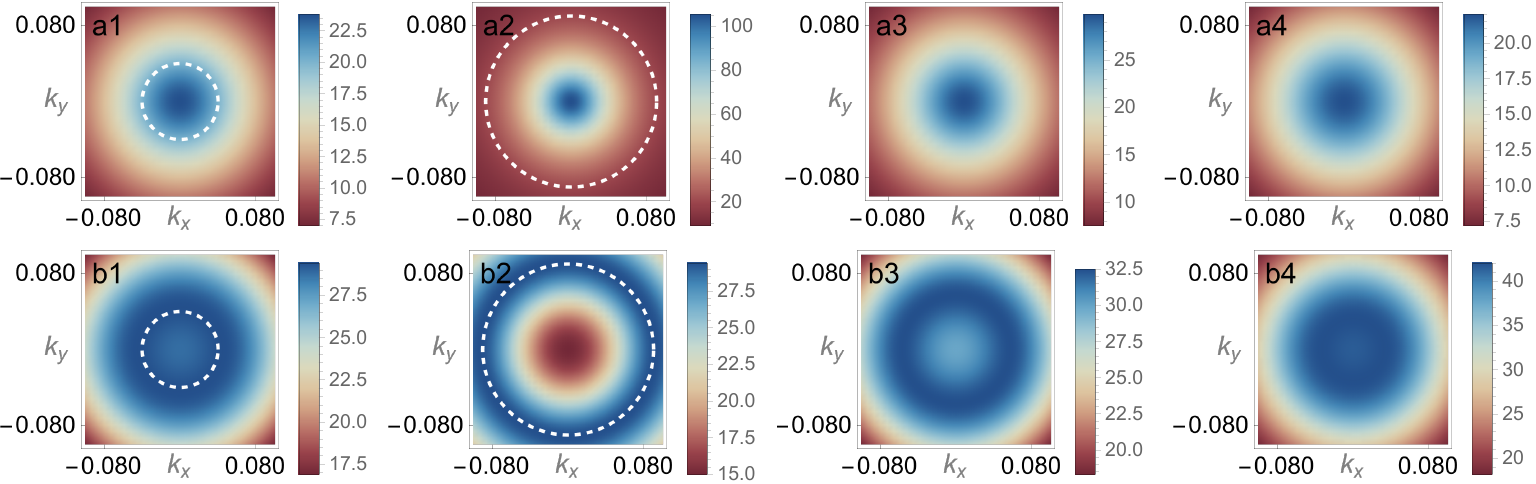}
\caption{Density channel signal profile of $\rho_0$ at zero temperature in the momentum $\bk$-space in the presence of $\gamma_3$-relaxation. Two energy planes at (a) $\varepsilon=0$ and (b) $\varepsilon=-0.06$ are shown. Both panels (a1-4) and (b1-4), from left to right, respectively use the parameters $(m,\gamma_3)=(0, 0.04), (0, 0.09), (0.01, 0.06), (0.05, 0.06)$ with $\chi=1,\gamma_0=0.1$ all along. White dashed line indicates the expected position of the ER encircling the bulk Fermi drumhead. Absence of white dashed line means the absence of ER.}\label{Fig:density_gamma3}
\end{figure}

On the other hand, we directly inspect the two specific cases so as to exemplify why the density channel provides much less nontrivial information about the NH system. 

\begin{figure}[hbt]
\includegraphics[width=16cm]{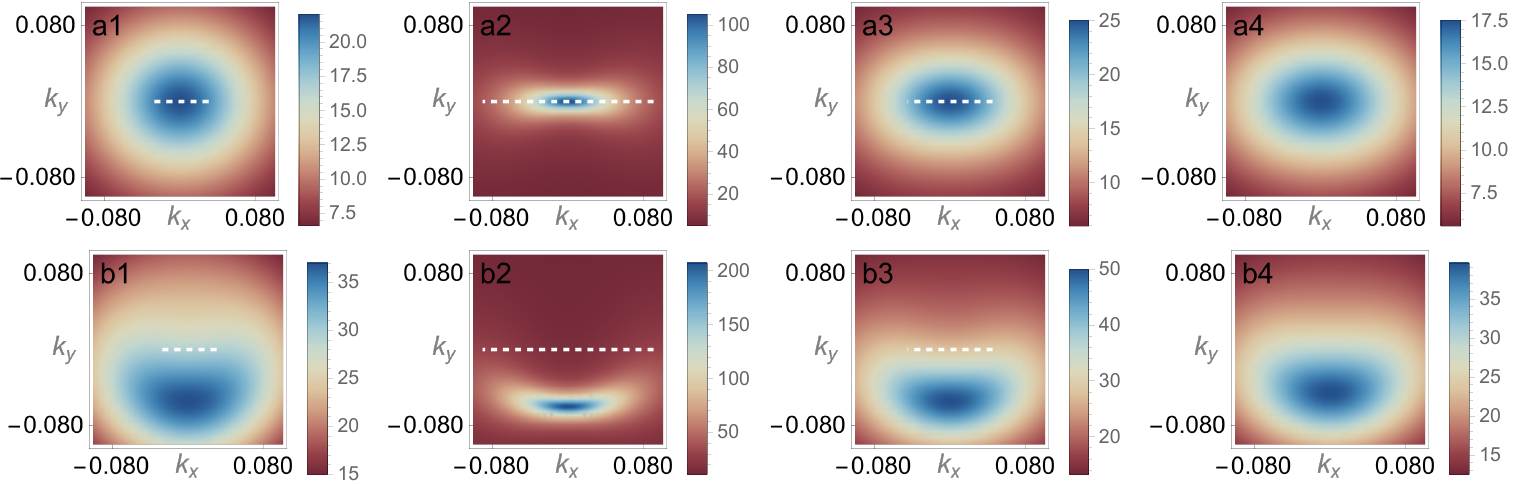}
\caption{Density channel signal profile of $\rho_0$ at zero temperature in the momentum $\bk$-space in the presence of $\gamma_1$-relaxation. Two energy planes at (a) $\varepsilon=0$ and (b) $\varepsilon=-0.06$ are shown. Both panels (a1-4) and (b1-4), from left to right, respectively use the parameters $(m,\gamma_1)=(0, 0.03), (0, 0.09), (0.05, 0.06), (0.07, 0.06)$ with $\chi=1,\gamma_0=0.1$ all along. White dashed line indicates the expected position of the bulk Fermi arc that ends at the two EPs. Absence of white dashed line means the absence of EPs.}\label{Fig:density_gamma1}
\end{figure}

\begin{itemize}
\item $\bgamma=\gamma_3\hat{z}$

We have
\begin{equation}\label{eq:rho_0_gamma3}
\begin{split}
    \rho_0(\gamma_3) & =  4 f(\omega) \frac{\gamma_0(\varepsilon^2+E_{\gamma_3}^2) + 2 \varepsilon m\gamma_3}{(\varepsilon^2-E_{\gamma_3}^2)^2 + 4(\varepsilon\gamma_0+m\gamma_3)^2} 
\end{split}
\end{equation}
with $E_{\gamma_3}^2=k^2+m^2+(\gamma_0^2-\gamma_3^2)$. As shown in Fig.~\ref{Fig:density_gamma3}, it is obvious that the profile is still circularly symmetric with respect to $\bk=0$. Given an energy $\varepsilon$-plane, the signal eventually decays at large momentum $k$ and $\gamma_3$ only quantitatively affects the strength: there exists no sharp disk-like feature in the 2D $\bk$-space. Note also that when $m=0$, i.e., the case with an ER as in Fig.~\ref{Fig:density_gamma3}(a1,a2), $\gamma_3$ only enters through $E_{\gamma_3}^2$ in a minor way: no distinguishing distinction between $\gamma_3=0$ and $\gamma_3>0$. The radius of the ER is also not clearly identifiable from the signal. Furthermore, there is no fundamental distinction between the cases with an ER, e.g., Fig.~\ref{Fig:density_gamma3}(a1,b1) and without an ER, e.g., Fig.~\ref{Fig:density_gamma3}(a3,b3). Even the extreme case with $\gamma_3$ very close to $\gamma_0$ only makes the nonmonotonic dependence of $k$ stronger in Fig.~\ref{Fig:density_gamma3}(b2), compared to either the other case with an ER in Fig.~\ref{Fig:density_gamma3}(b1) or the case without an ER in Fig.~\ref{Fig:density_gamma3}(b3). Hence, this nonmonotonic feature at finite energy plane is not a unique feature of ER, either. Here, the quantitative change in the profile is even less sharp than the $\gamma_1$ case since there is no symmetry change due to $\gamma_3$.
%This trend will be more drastic in the form of  when $\gamma_3>\gamma_0$, which is, however, unphysical for the present system as we have stressed.

\item $\bgamma=\gamma_1\hat{x}$ 
    
We have 
\begin{equation}
\begin{split}
    \rho_0(\gamma_1) & =  4 f(\omega) \frac{\gamma_0(\varepsilon^2+E_{\gamma_1}^2) - 2 \chi\varepsilon k_y \gamma_1}{(\varepsilon^2-E_{\gamma_1}^2)^2 + 4(\varepsilon\gamma_0-\chi k_y \gamma_1)^2} 
\end{split}
\end{equation}
with $E_{\gamma_1}^2=k^2+m^2+(\gamma_0^2-\gamma_1^2)$. As shown in Fig.~\ref{Fig:density_gamma1}, given a fixed energy $\varepsilon$-plane, the signal eventually decays at large momentum $k$ and the presence of $\gamma_1$ only breaks in a quantitative way the circular symmetry associated with the $\gamma_0$-only case. The profile satisfies the mirror symmetry with respect to $k_y$-axis, i.e., $\rho_0(k_x)=\rho_0(-k_x)$. However, the other mirror symmetry with respect to $k_x$-axis, i.e., $\rho_0(k_y)=\rho_0(-k_y)$, only holds in the $\varepsilon=0$ energy-plane. Hence, moving $\varepsilon$ can largely distort the profile as shown in Fig.~\ref{Fig:density_gamma1}(b), rendering the signal less robust. Moreover, the condition of forming a pair of EPs, i.e., $\gamma_1>|m|$, does not bring any sharp feature. For instance, within the $\varepsilon=0$ energy-plane, we have
\begin{equation}
\begin{split}
    \rho_0(\gamma_1,\varepsilon=0) & =  4 \gamma_0 f(\omega) \frac{E_{\gamma_1}^2}{E_{\gamma_1}^4 + 4 \gamma_1^2 k_y^2 } .
\end{split}
\end{equation}
The presence of $\gamma_1$ in $E_{\gamma_1}^2$ is minor; the second term $4 \gamma_1^2 k_y^2$ in the denominator makes a difference in the profile: the strength decays faster along $k_y$-axis than $k_x$-axis as shown in Fig.~\ref{Fig:density_gamma1}(a). In the extreme case when $m^2+(\gamma_0^2-\gamma_1^2)$ is very small, e.g., compared to $\gamma_1^2$, this asymmetry can become more obvious with a rod-like profile along $k_x$-axis as in Fig.~\ref{Fig:density_gamma1}(a2), because $E_{\gamma_1}^4$ is less dominant than $4 \gamma_1^2 k_y^2$. This is the closest of a signal akin to the bulk Fermi arc. However, from the comparison between Fig.~\ref{Fig:density_gamma1}(a1,a2,a3), the apparent rod length is not necessarily corresponding to the arc length, let along the foregoing $\varepsilon$-dependent distortion. To make it more susceptible, as $|m|$ or $(\gamma_0^2-\gamma_1^2)$ increases, such asymmetry is more and more obscured as the rod becomes akin to a rounder and rounder ellipse as in Fig.~\ref{Fig:density_gamma1}(a1,a3), which is a smooth deformation and continuous strength variation all along. Last but not the least, there is no fundamental difference between the cases with EPs in, e.g., Fig.~\ref{Fig:density_gamma1}(a3), and without EPs in Fig.~\ref{Fig:density_gamma1}(a4). In essence, the observable effect in the density channel is more of a quantitative circular symmetry breaking.

\end{itemize}

\subsection{Singularity in the spin texture}\label{SecSM:spin_channel}
Here, we derive the condition for having a singular point of the spin texture $\bp=0$ in terms of the orientation in the general case of Eq.~\eqref{eq:p_nu}, which is the most relevant to the topology. For such a vectorial property, it suffices to rotate the coordinate system such that $\bgamma$ is aligned to the $\hat{z}$-direction, leading to 
\begin{equation}\label{}
\begin{split}
    \bar\bp &= 2(\gamma_0\varepsilon+\bar d_3\gamma)\bar\bd + (\varepsilon^2-E^2) \gamma\hat{z}
\end{split}
\end{equation}
where $\bar\bd$ is rotated from $\bd$ and hence $E^2=\bar d^2+\gamma_0^2-\gamma^2$. Then $\bar\bp_{12}=0$ can be satisfied under the following two conditions. And we assume $\gamma_0>\gamma$ as discussed in the main text. Note that this constraint is crucial in determining which case can possess a singularity.
\begin{itemize}
\item $\varepsilon=-\frac{\gamma}{\gamma_0}\bar d_3$
    
Substituting this relation, we have 
\begin{equation}\label{}
\begin{split}
    \bar p_3= \gamma[(\frac{\gamma^2}{\gamma_0^2}-1)\bar d_3^2 - (\bar d_{12}^2+\gamma_0^2-\gamma^2) ]<0.
\end{split}
\end{equation}
This case will not present any singularity.
\item $\bar \bd_{12}=\bzero$

In this case, we can use the relation $\bar d_3^2=\bar d^2$ for $E^2$. Then we can obtain the following quadratic form with respect to $\varepsilon$
\begin{equation}\label{}
\begin{split}
    \bar p_3= \gamma\varepsilon^2 + 2\gamma_0 \bar d_3 \varepsilon + \gamma[\bar d^2-(\gamma_0^2-\gamma^2)] .
\end{split}
\end{equation}
And its discriminant 
\begin{equation}\label{}
\begin{split}
    \Delta&=  4\gamma_0^2 \bar d_3^2 - 4\gamma^2[\bar d^2-(\gamma_0^2-\gamma^2)] \\
    &= 4(\gamma_0^2-\gamma^2)(\bar d^2+\gamma^2)>0,
\end{split}
\end{equation}
which implies $\bar p_3=0$ is always possible. Let's relate it to the two specific cases in the main text. 
\begin{itemize}
\item $\bgamma=\gamma_1\hat{x}$ 
    
We actually have $\bar\bd=(-d_3,d_2,d_1)$. The condition $\bar \bd_{12}=0$ means $m=k_x=0$ and the solution is simply the two vortex cores $\bK^\pm$ in the vortex case, which is already discussed in the main text.
\item $\bgamma=\gamma_3\hat{z}$

We actually have $\bar\bd=\bd$. The condition $\bar \bd_{12}=0$ means $\bk=\bzero$. Then we obtain a meaningful solution of singularity at 
\begin{equation}\label{eq:varepsilon^pm}
    \varepsilon^s_\mathrm{mp}=\frac{1}{\gamma_3}\left[-m\gamma_0+s\sqrt{(\gamma_0^2-\gamma_3^2)(m^2+\gamma_3^2)}\right]
\end{equation}
for $s=\pm1$. 

Let's further consider the spin texture in the vicinity of the two singular points in the 3D $(\varepsilon,\bk)$-space. Substituting the position $(\varepsilon^\pm_\mathrm{mp}+\delta\varepsilon,\delta\bk)$ in Eq.~\eqref{eq:p_nu} and assuming $\delta\varepsilon,\delta k$ are arbitrarily small, we obtain the leading order expression
\begin{equation}\label{eq:deltap^pm}
\begin{split}
    \delta\bp^\pm=2(\gamma_0\varepsilon^\pm_\mathrm{mp}+\gamma_3m)\delta\bd_{12} + 2(\gamma_0m+\gamma_3\varepsilon^\pm_\mathrm{mp})\delta\varepsilon\hat{z}  
\end{split}
\end{equation}
where $\delta\bd_{12}=(-\chi\delta k_y,\delta k_x)$. This is indeed in a form that is linear in $\bdelta\equiv(\delta\varepsilon,\delta\bk)$, i.e., a hedgehog/monopole texture. A crucial property is that the coefficients satisfy
\begin{equation}\label{eq:deltap^pm_coeff0}
\begin{split}
    (\gamma_0\varepsilon^\pm_\mathrm{mp}+\gamma_3m)&= \frac{1}{\gamma_3}\left[\pm\gamma_0\sqrt{(\gamma_0^2-\gamma_3^2)(m^2+\gamma_3^2)} - m(\gamma_0^2-\gamma_3^2)\right] \gtrless 0 \\
     (\gamma_0m+\gamma_3\varepsilon^\pm_\mathrm{mp})&= \pm\sqrt{(\gamma_0^2-\gamma_3^2)(m^2+\gamma_3^2)}\gtrless0,
\end{split}
\end{equation}
where the first inequality for the less obvious $\varepsilon^+$ case can be shown as follows
\begin{equation}\label{}
\begin{split}
     \gamma_0^2(\gamma_0^2-\gamma_3^2)(m^2+\gamma_3^2) - m^2(\gamma_0^2-\gamma_3^2)^2= \gamma_3^2(\gamma_0^2-\gamma_3^2)(\gamma_0^2+m^2) >0 .
\end{split}
\end{equation}
Then we obtain for the form $\delta\bp^\pm=V^\pm\cdot\bdelta$ the sign of the coefficient tensor $V^\pm$ are 
\begin{equation}\label{eq:deltap^pm_coeff}
\begin{split}
\sgn (V^\pm)= \begin{bmatrix}
 &  & \mp\chi\\
 & \pm &  \\
\pm & & 
\end{bmatrix} .
\end{split}
\end{equation}
The skyrmion density from Eq.~\eqref{eq:deltap^pm} reads 
\begin{equation}\label{eq:b^pm}
\begin{split}
    \bb^\pm=\frac{1}{2}\epsilon^{ijk}\;\hat{i} \;\hat{\bp}^\pm\cdot(\partial_j \hat{\bp}^\pm\times\partial_k \hat{\bp}^\pm)
\end{split}
\end{equation}
where the vectorial indices are in terms of the $(\delta\varepsilon,\delta\bk)$-space.
The monopole charge is given by a spherical closed surface integral in the $(\delta\varepsilon,\delta\bk)$-space of the skyrmion density 
\begin{equation}\label{}
\begin{split}
    C^\pm=\oiint\dd\bS\cdot\bb^\pm=\pm\chi,
\end{split}
\end{equation}
which is consistent with another form $C^\pm=\sgn[\det(V^\pm)]$ easily calculated from Eq.~\eqref{eq:deltap^pm_coeff}.
\end{itemize}

\end{itemize}

Lastly, we inspect how the NH topological spin textures are created physically, for which we can focus on the position of the singularities (i.e., monopoles or meron/vortex cores) in the 3D $(\varepsilon,\bk)$-space.
\begin{itemize}
\item $\gamma_3$ case

We base on Eq.~\eqref{eq:varepsilon^pm}.
For the special massless case, we simply have $\varepsilon^\pm_\mathrm{mp}=\pm\sqrt{\gamma_0^2-\gamma_3^2}$. However, for the more generic massive case, as $\gamma_3\rightarrow0$ we obtain 
\begin{equation}\label{}
\begin{split}
    \varepsilon^\pm_\mathrm{mp}\rightarrow\frac{\gamma_0}{\gamma_3}\left(-m\pm|m|\right) \rightarrow \begin{cases}
    0,\,-\infty & m>0 \\
    \infty,\,0 & m<0 
    \end{cases}.
\end{split}
\end{equation}
As $\gamma_3$ increases from zero, for the massless case, the monopole pair is symmetrically created near $\varepsilon=\pm\gamma_0$; for the massive case, one in the pair at $\varepsilon_\mathrm{mp}^{\sgn(m)}$ is created near $\varepsilon=0$ while the other at $\varepsilon_\mathrm{mp}^{-\sgn(m)}$ is created from the energy infinity $-\sgn(m)\infty$. The mass term makes a difference in the pair creation and it partially follows from the fact that $\varepsilon=0$ is a singular plane in the $\gamma_0$ case.
The apparent discontinuity in the pair creation location between massless and massive cases merely reflects the fact that 
\begin{equation}
    \lim_{\gamma_3\rightarrow0}\lim_{m\rightarrow0}\varepsilon^s_\mathrm{mp}\neq\lim_{m\rightarrow0}\lim_{\gamma_3\rightarrow0}\varepsilon^s_\mathrm{mp}.
\end{equation}
From these properties, we can understand how the pair of monopole and antimonopole is created in general. 

Intuitively, the in-plane vorticity of the monopole-skyrmion texture is inherited from the original Hermitian spin-momentum locking of the topological insulator surface state. In the massless case, the symmetric monopole pair directly makes use of such vorticity to generate their associated skyrmions. However, in the massive case, one of the monopoles and its associated skyrmion make their way from the energy infinity in order to use such vorticity.

\item $\gamma_1$ case

We base on the momentum-space location of meron/vortex cores
\begin{equation}
\begin{split}
    \bK^\pm&=\left(0,\frac{1}{\gamma_1}\left[\chi\gamma_0\varepsilon\pm\sqrt{(\gamma_0^2-\gamma_1^2)(\varepsilon^2+\gamma_1^2)+\gamma_1^2m^2}\right]\right)
\end{split}
\end{equation}
given in the main text.
For the special $\varepsilon=0$-plane, we have $\bK^\pm=\left(0,\pm\sqrt{(\gamma_0^2-\gamma_1^2+m^2)}\right)$ symmetric with respect to the $k_x$-axis. On the other hand, for generic energy planes as $\gamma_1\rightarrow0$ we obtain 
\begin{equation}\label{}
\begin{split}
    K_y^\pm\rightarrow\frac{1}{\gamma_1}\left(\chi\gamma_0\varepsilon\pm|\gamma_0\varepsilon|\right) \rightarrow \begin{cases}
    0,\,-\infty & \sgn(\chi\varepsilon)=-1 \\
    \infty,\,0 & \sgn(\chi\varepsilon)=1 
    \end{cases}.
\end{split}
\end{equation}
As $\gamma_1$ increases from zero, for the special zero energy plane, the meron/vortex pair is symmetrically created near $\bk=(0,\pm\sqrt{\gamma_0^2+m^2})$; for the generic cases, one in the pair at $\bK^{-\sgn(\chi\varepsilon)}$ is created near $\varepsilon=0$ while the other at $\bK^{\sgn(\chi\varepsilon)}$ is created from the momentum infinity $k_y=\sgn(\chi\varepsilon)\infty$. 
The behavior resembles the $\gamma_3$ case in the sense that the special $\varepsilon=0$ case plays a similar role to the previous massless case.
The apparent discontinuity in the pair creation location between generic energy planes and the special zero energy plane similarly reflects the fact that 
\begin{equation}
    \lim_{\gamma_1\rightarrow0}\lim_{\varepsilon\rightarrow0}\bK^s\neq\lim_{\varepsilon\rightarrow0}\lim_{\gamma_1\rightarrow0}\bK^s.
\end{equation}
From these properties, we can understand how the pair of merons/vortices is created in general. 

Intuitively, the in-plane vorticity of the meron/vortex pair texture is again inherited from the original Hermitian surface state spin-momentum locking. However, it is very intriguing that NH relaxations can make a qualitative difference: such vorticity is used twice to create the new topological soliton pair structure.
For generic energy planes, such a procedure is made possible at the momentum-space infinity where the extra meron/vortex is created, in contrast to the energy infinity in the $\gamma_3$ case.
\end{itemize}

\section{Spin textures due to general direction of NH dissipation}\label{Sec:theta}

In the main text and previous discussions, we focus on two cases of dissipation or NH effect, $\bgamma=\gamma_3\hat{z}$ and $\bgamma=\gamma_1\hat{x}$. They correspond to physically representative and accessible situations: the dissipation vector is normal or in-plane to the surface. They can in general be obtained from the coupling to a bath as formulated in Sec.~\ref{SecSM:bath}. For the typical relaxation mechanism of magnetic impurity scattering, discussed in the introduction part of the main text and also in Sec.~\ref{SecSM:imp}, the two cases, in the same manner, respectively correspond to magnetic doping that is normal or in-plane to the sample surface, which are readily achievable in experiments of topological insulator. Hence, it is naturally expected that they are epitomes of the proposed phenomenon. 

However, it is as well physically intriguing what happens with intermediate cases, i.e., those with general direction of $\bgamma$ vector. Experimentally, this is also very useful as it helps verification in more general settings and relaxes the requirement. 
We consider in detail below the generic 
\begin{equation}
    \bgamma=\gamma\sin{\theta}\hat{x}+\gamma\cos{\theta}\hat{z}=\gamma_1\hat{x}+\gamma_3\hat{z} \qquad \theta\in[0,\pi/2].
\end{equation}
Unless otherwise stated, we restrict ourselves to this parameter setting. Firstly, the case with both $\gamma_1$ and $\gamma_2$ present is actually not different from the pure $\gamma_1$ case up to a trivial rotation. Secondly, other directions can be simply related to this via symmetries. 
Although not limited to the scenario of dissipation or non-Hermitian terms from magnetization induced scattering, we can use it as an example. Given a generic surface magnetization due to magnetic doping
\begin{equation}
    \bmm=m\sin{\theta}\hat{x}+m\cos{\theta}\hat{z},
\end{equation}
our model Eq.~\eqref{eq:H0} would become
\begin{equation}\label{eq:H0_M}
    H_0=k_x\sigma_2 - \chi (k_y-\chi m_x)\sigma_1 + m_z\sigma_3
\end{equation}
with $m_x=m\sin{\theta},m_z=m\cos{\theta}$. The effect of $m_x$, as mentioned in the main text, is merely shifting the spin texture profile in $k_y$; $m_z$ leads to the Dirac mass denoted as $d_3$ or $m$ throughout the discussion for brevity. Intrinsic magnetism in the system may add to the magnetization without causing much impurity scattering relaxation. Therefore, in the following, without loss of generality, we can keep using the representative Eq.~\eqref{eq:H0} with the shift from in-plane magnetization understood.

To facilitate the discussion, we perform a rotation on the spin texture Eq.~\eqref{eq:p_nu_maintext} such that $\bgamma$ is aligned in the $\hat{x}$-direction
\begin{equation}\label{eq:barp}
    \bar\bp=R\cdot\bp=2(\gamma_0\varepsilon+\bar d_1\gamma)\bar\bd + (\varepsilon^2-E^2)\gamma\hat{x}
\end{equation}
with $E^2=d^2+\gamma_0^2-\gamma^2$, where $\bar\bd=R\cdot\bd$ is defined in the same manner as $\bar\bp$ and 
\begin{equation}\label{eq:R_rot}
    R=\begin{bmatrix}
    \sin{\theta} & 0 & \cos{\theta} \\
    0 & 1 & 0 \\
    -\cos{\theta} & 0 & \sin{\theta} \\
    \end{bmatrix}.
\end{equation}
We henceforth base the discussion on this texture without loss of generality. This is because the transformation only globally rotates the spin space without touching any of the topological properties.
In addition, we henceforth denote and use $\tilde k_y=\chi k_y$ and $\tilde\bk=(k_x,\tilde k_y)$ for the brevity of discussion. 

Before the detailed analysis, we summarize the most important messages and new features here.
\begin{itemize}
    \item The main crucial features we highlight in the main text, i.e., the monopole pair and the vortex pair, mostly persist and coexist for generic $\theta$ and are smoothly connected to the extreme cases of pure $\gamma_3$ ($\theta=0$) or $\gamma_1$ ($\theta=\pi/2$). The feature of monopole pair highly resembles the pure $\gamma_3$ case.
    \item Entirely new features associate the vortex formation, including a critical angle $\theta=\pi/4$ and topological transitions between vortex-vortex pair  and vortex-antivortex pair . %See also the summary above Sec.~\ref{Sec:theta>45}.
\begin{itemize}
    \item For any $\varepsilon$-planes one has a vortex-vortex pair if and only if $\pi/4<\theta\leq\pi/2$. This highly resembles the pure $\gamma_1$ case of $\theta=\pi/2$.
    \item $\theta=\pi/4$ is singled out as a critical angle with only one vortex. It is a topological transition point of the system's overall behavior.
    \item When $\theta<\pi/4$, one has a vortex-antivortex pair only for some $\varepsilon$-planes satisfying $\varepsilon\gtrless\varepsilon_\mathrm{vt}^\pm$. Moving inside the two critical $\varepsilon_\mathrm{vt}^\pm$ planes, the pair annihilates as another distinct topological transition.
\end{itemize}
\end{itemize}
Similar to the discussion in the main text, these will be distinct and extra experimentally detectable phenomena.

\subsection{Monopole-antimonopole pair}\label{Sec:theta_mp}

We first consider the possibility of monopole singularity in the spin texture of Eq.~\eqref{eq:barp}.
\begin{itemize}
\item $\varepsilon=-\frac{\gamma}{\gamma_0}\bar d_1$
    
We have $\bar\bp_{23}=0$ under this condition. Substituting this, we find that 
\begin{equation}\label{eq:barp_1<0}
    \bar p_1=\gamma[(\frac{\gamma^2}{\gamma_0^2}-1)\bar d_1^2 - (\bar d_{23}^2+\gamma_0^2-\gamma^2) ]<0.
\end{equation}
This case will not present any singularity except a plane in the 3D $\bX=(\varepsilon,\tilde\bk)$ space where $\bar p_1<0,\bar\bp_{23}=0$.

\item $\bar\bd_{23}=0$

We automatically have $\bar\bp_{23}=0$ under this condition and 
\begin{equation}\label{eq:epsilon_mp}
    \bar p_1(\varepsilon)=\gamma\varepsilon^2+2\gamma_0d_3\sec{\theta}+\gamma(\gamma_0^2-\gamma^2+d_3^2\sec^2{\theta})=0
\end{equation}
has solution for monopole when 
\begin{equation}
    \varepsilon_\mathrm{mp}^s = \frac{-\gamma_0d_3\sec{\theta}+s\frac{\sqrt{\Delta_\mathrm{mp}}}{2}}{\gamma}
\end{equation}
with $s=\pm1$ and the positive-definite discriminant
\begin{equation}
    \Delta_\mathrm{mp} = 4(\gamma_0^2-\gamma^2)(d_3^2\sec^2{\theta}+\gamma^2).
\end{equation}

\end{itemize}

Thus, we obtain two singular monopolar points $\bX_\mathrm{mp}^\pm=(\varepsilon_\mathrm{mp}^\pm,\tilde\bk_\mathrm{mp})$ for almost \textit{any} $\theta$ with $\tilde\bk_\mathrm{mp}=(0,-d_3\tan\theta)$. Note that the monopole location is in general shifted in the $\tilde k_y$-axis due to nonzero $\theta$ and the special massless case $d_3=0$ remains the same as the pure $\gamma_3$ case. The only exception lies in the case when $\theta$ approaches $\frac{\pi}{2}$: we have for both $s=\pm1$
\begin{equation}\label{eq:epsilon_mp_lim}
\begin{split}
    &\lim_{\theta\rightarrow\frac{\pi}{2}^-} \varepsilon_\mathrm{mp}^s= \sgn[s(\gamma_0^2-\gamma^2)d_3^2-\gamma_0^2d_3^2]\,\infty \\
    &s(\gamma_0^2-\gamma^2)d_3^2-\gamma_0^2d_3^2 \lessgtr0 \qquad d_3\gtrless0.
\end{split}
\end{equation}
This implies that the monopole pair will move to the positive or negative energy plane at infinity, dependent on the sign of Dirac mass $d_3=m$. In reality, they will eventually merge into the bulk bands. This disappearance of the monopole pair when $\theta=\pi/2$ is fully expected as it is exactly the pure $\gamma_1$ case discussed in the main text.

Now we inspect the spin texture in the vicinity of the monopoles, for which we find the form up to the leading order
\begin{equation}
    \bar\bp(\bX^s+\delta\bX)=V^s\cdot\delta\bX+O(\delta \bX^2)
\end{equation}
where $\delta\bX$ is measured from $\bX_\mathrm{mp}^s$ and
\begin{equation}
    V^s=\begin{bmatrix}
    s\sqrt{\Delta_\mathrm{mp}} & 0 & -\sin{\theta}\,a_s \\
    0 & a_s & 0 \\
    0 & 0 & \cos{\theta}\,a^s
    \end{bmatrix}
\end{equation}
with $a_s=[s\gamma_0\sqrt{\Delta_\mathrm{mp}}-2(\gamma_0^2-\gamma^2)d_3\sec{\theta}]/\gamma$. One can immediately calculate $\Det V^s=s\cos{\theta}\sqrt{\Delta_\mathrm{mp}}a_s^2$ and see the topological charge
\begin{equation}
C^s=\chi\sgn[\Det V^s]=\chi s    
\end{equation}
for $\theta\in[0,\pi/2)$ of our focus. Note that here we include the $\chi$-dependence as it is given in terms of the $(\varepsilon,\bk)$ space. This is consistent with the pure $\gamma_3$ case discussed in the main text and in Sec.~\ref{SecSM:spin_channel}. The monopole textures will be qualitatively similar to those in Fig.~\ref{Fig:skyrmion} and hence not repeated here.
It is also important to note that the skyrmion number $N_\mathrm{sk}$ defined in each $\varepsilon$-plane will always exhibit a \textit{quantized} jump of $\pm1$ when we vary $\varepsilon$ and cross the monopole or antimonopole points. The only difference at generic $\theta$ is that the skyrmion number for arbitrary $\varepsilon$-plane is not necessarily quantized since the defining manifold is not necessarily closed due to the nonuniform boundary condition at the infinity. It remains a constant value until the monopole or antimonopole is reached and the foregoing quantized jump occurs. Therefore, the appearance of a monopole-antimonopole pair in the energy-momentum space and the quantized jump of topological charge is proved to be a stable feature due to the non-Hermitian effects, which is experimentally detectable via a tomography of the spin texture as we proposed. 

\subsubsection{Relation to the vortex features}
%\subsubsection{Relation to the topological transition points of vortex-antivortex annihilation}
In this subsection, previewing the discussion of vortex pair in Sec.~\ref{Sec:theta_vt} when necessary, we discuss the relation between the two apparently distinct features of the monopole pair and the vortex pair. Firstly, they are not mutually exclusive for generic $\theta$, except for the two boundary cases expounded in the main text, i.e, the pure $\gamma_3$ case when $\theta=0$ and the pure $\gamma_1$ case when $\theta=\pi/2$. When $\theta=\pi/2$, we already mentioned around Eq.~\eqref{eq:epsilon_mp_lim} that the monopole pair will disappear at the infinity of energy and merge into bulk bands. When $\theta=0$, as is to be mentioned shortly below, the vortex pair solution ceases to be meaningful and only the monopole pair and skyrmion are left. Within the wide region in between, i.e., $0<\theta<\pi/2$, there is no contradiction in the presence and condition of the monopole pair and the vortex pair: they will simply \textit{coexist}. 

On the other hand, it is still very helpful to see how some features of special energy planes are smoothly connected from the vortex context to the monopole context, which is especially related to the $\theta<\pi/4$ case in Sec.~\ref{Sec:theta<45}. We know that there exists a pair of critical energy planes $\varepsilon_\mathrm{vt}^t$ given in Eq.~\eqref{eq:epsilon_vt} when $\theta<\pi/4$, where a pair of vortex and antivortex annihilates at $\tilde\bk^t$ given in Eq.~\eqref{eq:tildebk^t}. We can compare it with the monopole energy planes Eq.~\eqref{eq:epsilon_mp} and find that (proof given after the discussion) when $0<|\theta|<\pi/4$
\begin{equation}\label{eq:epsilon_mp_vt}
    \varepsilon_\mathrm{mp}^-<\varepsilon_\mathrm{vt}^- < \varepsilon_\mathrm{vt}^+<\varepsilon_\mathrm{mp}^+
\end{equation}
and 
\begin{equation}
    \varepsilon_\mathrm{mp}^s(\theta=0)=\varepsilon_\mathrm{vt}^{t=s}(\theta=0).
\end{equation}
In addition, in the momentum space, we notice that $\tilde\bk^t(\theta=0)=\bzero$ coincides with the foregoing position of the monopoles $\tilde\bk_\mathrm{mp}(\theta=0)=\bzero$.
This means that the two vortex pair annihilation energy planes always lie in between the monopole planes; the two vortex pair annihilation points in the 3D $\bX=(\varepsilon,\tilde\bk)$ space \textit{exactly} approach to the monopole pair at $\theta=0$, i.e., the pure $\gamma_3$ case. We comment from two aspects. Firstly, the topological vortex annihilation points, which are mainly for $\theta\neq0$, exactly deform to our monopole pair in the $\gamma_3$ case in the main text and it is thus featureless between the pair as expected. Secondly, outside the monopole pair there seems to remain vortex-antivortex pair even at $\theta=0$. However, this is merely an artifact of the vortex analysis based on the rotated spin space: for our pure $\gamma_3$ case the rotation in Eq.~\eqref{eq:R_rot} is at the extreme case of $\theta=0$, which is, equivalently speaking, a $\pi/2$-rotation with respect to $\hat{y}$. For each Bloch-type skyrmion plane, it thus identifies the vicinity around two locations where the spin points in the $\pm\hat{x}$ direction as votices, which is unnecessary due to the circular symmetry at exactly $\theta=0$ in each energy plane. 
Therefore, in summary, the analysis in Sec.~\ref{Sec:theta_vt} smoothly connects to our general monopole analysis herein and especially the pure $\gamma_3$ case in the main text in a fully reasonable manner.

To see Eq.~\eqref{eq:epsilon_mp_vt}, we calculate
\begin{equation}\label{eq:epsilon_diff_mpvt}
    \varepsilon_\mathrm{mp}^s-\varepsilon_\mathrm{vt}^s=\frac{a+s(b-c)}{\gamma Q}
\end{equation}
where 
\begin{equation}
\begin{split}
    a= -\gamma_0d_3\sec{\theta}\sin^2{\theta}(\gamma_0^2-\gamma^2),\quad
    b= \frac{Q}{2}\sqrt{\Delta_\mathrm{mp}},\quad
    c=\frac{\gamma^2}{8}\sqrt{\Delta_\mathrm{vt}}
\end{split}
\end{equation}
with $Q$ and $\Delta_\mathrm{vt}$ respectively given by Eq.~\eqref{eq:Q_def} and Eq.~\eqref{eq:Delta_vt}. One can prove in three steps to see that 
\begin{equation}\label{eq:b-c}
    b-c>|a|
\end{equation}
as follows. Firstly, we have $b>c>0$ since
\begin{equation}
\begin{split}
    b^2-c^2&=(\gamma_0^2-\gamma^2)[Q^2(\gamma^2+d_3^2\sec^2{\theta})-\gamma^4\cos{2\theta}(Q+d_3^2)] \\
    &=(\gamma_0^2-\gamma^2)[\gamma^2Q(Q-\gamma^2\cos{2\theta})+d_3^2\sec^2{\theta}(Q^2-\gamma^4\cos^2{\theta}\cos{2\theta})]\\
    &=(\gamma_0^2-\gamma^2)[\gamma^2Q(W+\gamma^2\sin^2{\theta})+d_3^2\sec^2{\theta}(W^2+\gamma^4\cos^2{\theta}\sin^2{\theta}+2W\gamma^2\cos^2{\theta})]>0
\end{split}
\end{equation}
with $W=\sin^2{\theta}(\gamma_0^2-\gamma^2)$. 
Secondly, we have 
\begin{equation}
\begin{split}
    b^2+c^2-a^2
    &= \frac{1}{8} \gamma^2 (\gamma_0^2-\gamma^2) Q \sec ^2\theta [(4\gamma^2-\gamma_0^2) \cos 4 \theta +\gamma_0^2+4 (2\gamma ^2+3 d_3^2) \cos 2 \theta +4 (\gamma ^2+d_3^2)] \\
    &=\frac{1}{4}\gamma^2(\gamma_0^2-\gamma^2)Q\sec^2{\theta}[2(1+3\cos{2\theta})d_3^2+\gamma^2(2\cos^22\theta+4\cos2\theta+1)+Q|_{\theta\rightarrow2\theta}]>0
\end{split}
\end{equation}
since the first two parts inside the brackets are positive when $|\theta|<\pi/4$. Thirdly, we have \begin{equation}
    (b^2+c^2-a^2)^2-4 b^2 c^2 = \gamma^4(\gamma_0^2-\gamma^2)^2Q^2\tan^4{\theta}(d_3^2+\gamma_0\cos^2\theta)^2>0.
\end{equation}
Combining the above three inequalities, we can see $(b-c)^2-a^2>0$ and further the relation Eq.~\eqref{eq:b-c}. Together with Eq.~\eqref{eq:epsilon_diff_mpvt}, we know that $a\pm (b-c)\gtrless0$, which implies the key relation Eq.~\eqref{eq:epsilon_mp_vt}.

\subsection{Vortex pair and topological transitions}\label{Sec:theta_vt}

Now we consider the possibility of vortex formation in Eq.~\eqref{eq:barp}. The plane of $\bar\bp_{23}=0$ in the 3D $\bX$-space mentioned around Eq.~\eqref{eq:barp_1<0} in Sec.~\ref{Sec:theta_mp} obviously does not correspond to vortices. Following the $\gamma_1$ case in the main text, we consider $\bar\bp_{12}=0$. As $\bar d_2=d_2=0$ guarantees $\bar p_2=0$, we inspect 
\begin{equation}\label{eq:barp1_d1}
    \bar p_1(d_1)=-\gamma\cos{2\theta} \,d_1^2 + 2\sin\theta(\gamma_0\varepsilon+2\gamma d_3\cos\theta)d_1 + \{\gamma\cos2\theta \,d_3^2 + 2\gamma_0\cos\theta\,\varepsilon d_3+\gamma[\varepsilon^2-(\gamma_0^2-\gamma^2)]\}=0 ,
\end{equation}
whose discriminant is 
\begin{equation}\label{eq:Delta_barp1}
    \Delta\equiv \Delta_{d_1}^{\bar p_1}=4[(\gamma d_3+\cos\theta\,\gamma_0\varepsilon)^2-\cos{2\theta}(\gamma_0^2-\gamma^2)(\varepsilon^2+\gamma^2)]
\end{equation}
where we introduce the notation $\Delta_{x}^{f}$ meaning the discriminant of a quadratic equation $f(x)=0$. In order to have $\Delta\geq0$, we look at its own discriminant 
\begin{equation}\label{eq:Delta_vt}
    \Delta_\mathrm{vt}(\varepsilon)\equiv \Delta_\varepsilon^\Delta=64\gamma^2\cos{2\theta}(\gamma_0^2-\gamma^2)(d_3^2+Q)
\end{equation}
with 
\begin{equation}\label{eq:Q_def}
    Q=(\gamma_0^2-\gamma^2)\sin^2{\theta}(\gamma_0^2-\gamma^2)+\gamma^2\cos^2{\theta}>0.
\end{equation}
Given the quadratic coefficient of $\Delta_\mathrm{vt}(\varepsilon)$ is $4Q>0$, we see that $\Delta\geq0$ when $\Delta_\mathrm{vt}\leq0$, which implies the key dependence of vortex formation on $\theta$
\begin{itemize}
    \item For any $\varepsilon$-planes $\bar p_1=0$ has solution(s) for vortex if and only if $\pi/4\leq\theta\leq\pi/2$;
    \item When $\theta<\pi/4$, $\bar p_1=0$ has solution(s) for vortex only for some $\varepsilon$-planes satisfying $\Delta>0$.
\end{itemize}
This is our first key result and new feature due to finite $\theta$ in terms of vortex formation, which singles out the \textit{critical angle} $\theta=\pi/4$.
In the following, we will successively study the three cases based on the observation above. Note that a summary is given below Eq.~\eqref{eq:R_rot}. 
% \begin{itemize}
%     \item For any $\varepsilon$-planes one has a vortex-vortex pair if and only if $\pi/4<\theta\leq\pi/2$. This highly resembles the pure $\gamma_1$ case of $\theta=\pi/2$.
%     \item $\theta=\pi/4$ is singled out as a critical angle with only one vortex. It is a topological transition point of the system's overall feature.
%     \item When $\theta<\pi/4$, one has a vortex-antivortex pair only for some $\varepsilon$-planes satisfying $\varepsilon\gtrless\varepsilon_\mathrm{vt}^\pm$. Moving inside the two critical $\varepsilon_\mathrm{vt}^\pm$ planes, the pair annihilates as another distinct topological transition.
% \end{itemize}

\subsubsection{\texorpdfstring{$\theta>\pi/4$ case}{>45}}\label{Sec:theta>45}
\begin{figure}[hbt]
\includegraphics[width=17.8cm]{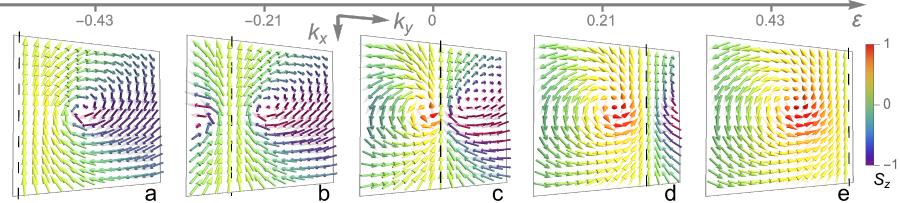}
\caption{Topological spin textures in the $(\varepsilon,\bk)$-space with $\bgamma$-relaxation in the direction $\theta=3\pi/8$ measured from $\hat{z}$ towards $\hat{x}$. Spins are in the globally rotated frame to facilitate presentation. (a,b,c,d,e) Five energy planes in the square-shaped $\bk$-space $[-0.8,0.8]^2$. A vortex-vortex pair appears in each energy plane in the similar manner of Fig.~\ref{Fig:meron} and the discussion of pure $\gamma_1$ case in the main text. Dashed black line indicates the borderline between the pair where spins are uniformly aligned. As the borderline shifts with $\varepsilon$, one vortex moves outside the plot region in (a,e). Parameters are $\chi=1,m=0.13,\gamma_0=0.1,\gamma_1=0.06$.}\label{Fig:theta>45}
\end{figure}

We first obtain some results for generic $\theta$ as long as $\Delta>0$.

According to Eqs.~\eqref{eq:barp1_d1}\eqref{eq:Delta_barp1}, the vortex cores are at $\bd_{12}^s=(d_1^s,0)$ with $s=\pm1$ and
\begin{equation}\label{eq:d1^s}
    d_1^s = \frac{1}{-\gamma\cos2\theta}\left[-\sin\theta(\gamma_0\varepsilon+2\gamma d_3\cos\theta)+\frac{s\sqrt{\Delta}}{2}\right].
\end{equation}
However, it is crucial to note $d_1^+\lessgtr d_1^-$ when $\cos2\theta\gtrless0$, i.e, when $\theta\lessgtr\pi/4$. Then the alternative expression of the location of vortex cores is $\tilde\bK^s=(0,\tilde k_y^s)$ with
\begin{equation}\label{eq:ky^s}
    \tilde K_y^s = \frac{1}{\gamma\cos2\theta}\left[-\sin\theta(\gamma_0\varepsilon+2\gamma d_3\cos\theta)-\sgn(\theta-\frac{\pi}{4})\frac{s\sqrt{\Delta}}{2}\right],
\end{equation}
which assures the convention $\tilde K_y^+>\tilde K_y^-$. 
Using variables $d_1,d_2$ and Eq.~\eqref{eq:d1^s} helps to obtain the general form in the vicinity of cores with the substitution $d_1= d_1^s+\delta d_1$
\begin{equation}\label{eq:barp12^s}
\begin{split}
    \bar \bp_{12}^s(\delta d_1,d_2)=\left(
    s\sqrt{\Delta}\delta d_1 - \gamma\cos2\theta \,\delta d_1^2 - \gamma d_2^2,\: 
    2\sec2\theta\,(\alpha'-s\,g)d_2+2\gamma\sin\theta\,\delta d_1d_2
    \right)
\end{split}
\end{equation}
where $g=\sin\theta\,\sqrt{\Delta}/2$ and 
\begin{equation}\label{eq:alpha_def}
    \alpha'=\alpha\cos\theta,\quad\alpha(\varepsilon,\theta)=\gamma d_3+\cos\theta\,\gamma_0\varepsilon.
\end{equation}
Due to finite $\theta$, $\bar p_3$ does not have a fixed sign in general and the exact half integral quantization of meron does not hold as in the pure $\gamma_1$ case mentioned in the main text.
Along the special line in each energy plane satisfying $\gamma_0\varepsilon+\gamma\bar d_1=0$, i.e, 
\begin{equation}\label{eq:tildeKy^0}
    \tilde K_y^0=\frac{\gamma_0 \varepsilon+\cos\theta\,\gamma d_3}{\gamma\sin\theta},
\end{equation}
the spins point all in the $-\hat{x}$ direction since $\bar\bp=\bar p_1\hat{x}$ and we have $\bar p_1=\gamma(\varepsilon^2-E^2)<0$. 
One also finds that
\begin{equation}\label{eq:borderline_diff}
\begin{split}
    \tilde K_y^s - \tilde K_y^0 &=- \frac{\alpha'+\sgn(\theta-\frac{\pi}{4})\,s\,g}{\gamma\sin\theta\cos2\theta} \\
    \alpha'^2-g^2&=\cos2\theta[\alpha^2+\sin^2\theta(\gamma_0^2-\gamma^2)(\varepsilon^2+\gamma^2)].   
\end{split}
\end{equation}
An important relation between vortex cores and this special spin-unidirectional line can be seen in the following.

Now we stick to $\pi/4<\theta<\pi/2$. Eq.~\eqref{eq:borderline_diff} implies that $|\alpha'|<g$ and hence
\begin{equation}\label{eq:borderline_relation_>45}
    \tilde K_y^-<\tilde K_y^0<\tilde K_y^+,
\end{equation}
i.e., the special spin-unidirectional line Eq.~\eqref{eq:tildeKy^0} always serves as the borderline in between the vortex-vortex pair, which is the same as the pure $\gamma_1$ case in the main text. However, we note that this borderline in general is not midway between the vortices. To see this, we find that
\begin{equation}
    \tilde K_y^0-\frac{\tilde K_y^++\tilde K_y^-}{2} = \frac{\cot\theta}{\cos2\theta}\alpha(\varepsilon,\theta),
\end{equation}
which vanishes, e.g., when $\cos\theta=0$. Hence, the pure $\gamma_1$ case is a special case where the borderline is indeed midway as mentioned in the main text.
According to Eq.~\eqref{eq:barp12^s}, up to the leading order near the core, we have
\begin{equation}
    \bar\bp_{12}(\tilde\bK^s)=(s\sqrt{\Delta}\delta\tilde k_y,2\sec2\theta\,(\alpha'-s\,g)k_x)\sim(s\delta\tilde k_y,-sk_x),
\end{equation}
where the last expression leaves all but the sign information, which gives a \textit{vortex-vortex pair} characterized by
\begin{equation}\label{eq:vt_character_>45}
    \mathtt{v}=\chi,\mathtt{h}=-s\frac{\pi}{2}.
\end{equation}
Since the vortex pair feature always persists, Fig.~\ref{Fig:theta>45} illustrates the $\varepsilon$-dependent configuration of the vortex pair in close analogy to Fig.~\ref{Fig:meron}. Also, as the generalization of the in-plane $d$-wave-like texture $\bp_\mathrm{as}$ in the main text, the asymptotic behavior at large momentum reads 
\begin{equation}
    \bar\bp_\mathrm{as} = -\gamma(\cos2\theta\,k_y^2+k_x^2,\:2\chi\sin\theta k_xk_y,\:\sin2\theta k_y^2),
\end{equation}
the in-plane part of which again exhibits a topological winding number $2\chi$ for generic $\theta>\pi/4$. 

\begin{itemize}
    \item In summary, the formation of a vortex-vortex pair as the main feature stably persists within the wide range $\pi/4<\theta\leq\pi/2$, where $\theta=\pi/2$ is the pure $\gamma_1$ case in the main text. 
\end{itemize}

\subsubsection{\texorpdfstring{$\theta=\pi/4$ case}{=45}}\label{Sec:theta=45}

At this critical angle, Eq.~\eqref{eq:barp1_d1}, which is quadratic otherwise, becomes linear. Hence, we have \textit{only one} solution for vortex core
\begin{equation}
    \tilde K_y^*(\varepsilon)=-\frac{\sqrt{2}\gamma_0\varepsilon d_3+\gamma[\varepsilon^2-(\gamma_0^2-\gamma^2)]}{\sqrt{2}(\gamma_0\varepsilon+\sqrt{2}\gamma d_3)}.
\end{equation}
Let us check its relation to the special spin-unidirectional line Eq.~\eqref{eq:tildeKy^0}
\begin{equation}
\begin{split}
    \tilde K_y^* - \tilde K_y^0 &= Nm(\varepsilon)/Dn(\varepsilon) \\
    Nm(\varepsilon) &= (2\gamma_0^2-\gamma^2)\varepsilon^2+ 2\sqrt{2}\gamma_0\gamma d_3\varepsilon+\gamma^2(2d_3^2+\gamma_0^2-\gamma^2) \\
    Dn(\varepsilon) &=-\sqrt{2}\gamma(\gamma_0\varepsilon+\sqrt{2}\gamma d_3).
\end{split}
\end{equation}
One finds the discriminant of the numerator
\begin{equation}
    \Delta_\varepsilon^{Nm} = -4\gamma^2(\gamma_0^2-\gamma^2)(2d_3^2+2\gamma_0^2-\gamma^2)<0,
\end{equation}
i.e., $Nm(\varepsilon)>0$ for any $\varepsilon$. Also, the denominator satisfies $\sgn(Dn)=-\sgn(\alpha)$ when $\theta=\frac{\pi}{4}$ with $\alpha$ given by Eq.~\eqref{eq:alpha_def}. This leads to 
\begin{equation}
    -\sgn(\tilde K_y^* - \tilde K_y^0)=\sgn[\alpha(\varepsilon,\theta=\frac{\pi}{4})]=\sgn[\varepsilon-\varepsilon^0(\theta=\frac{\pi}{4})].
\end{equation}
Therefore, the single vortex lies below (above) the spin-unidirectional line along the $\tilde k_y$-axis when the energy plane $\varepsilon\gtrless\varepsilon^0$. This is shown in Fig.~\ref{Fig:theta=45}. Here and also for later use, we introduce a special energy plane for generic $\theta\leq\frac{\pi}{4}$ 
\begin{equation}\label{eq:epsilon^0_def}
    \varepsilon^0=-\gamma d_3/(\gamma_0\cos\theta),
\end{equation}
which makes $\alpha(\varepsilon^0)=0$. 

In the vicinity of the vortex core at $\tilde\bK^*=(0,\tilde K_y^*)$, the in-plane spin texture reads from Eq.~\eqref{eq:barp}
\begin{equation}
    \bar\bp_{12}^* = \left(Dn\delta\tilde k_y/\gamma - \gamma k_x^2,\: -\sqrt{2}\gamma(\tilde K_y^* - \tilde K_y^0)k_x-\sqrt{2}\gamma\delta\tilde k_yk_x\right)\sim \sgn[\alpha(\varepsilon,\theta=\frac{\pi}{4})](-\delta k_y,k_x),
\end{equation}
where the last expression leaves all but the sign information. It gives a \textit{single vortex} characterized by
\begin{equation}\label{eq:vt_character_=45}
    \mathtt{v}=\chi,\mathtt{h}=\sgn(\alpha)\frac{\pi}{2}.
\end{equation}
We also notice that the above $\tilde K_y^*(\varepsilon)$ approaches $-\sgn(\varepsilon-\varepsilon^0)\infty$ when $\varepsilon$ moves towards $\varepsilon^0$, which is how the opposite-helicity vortices below or above $\varepsilon^0$ are connected across the special energy plane with no vortex.
To understand the appearance of this single vortex, we can check the situation when $\theta\rightarrow\frac{\pi}{4}^+$ based on Eqs.~\eqref{eq:ky^s}\eqref{eq:borderline_relation_>45}\eqref{eq:vt_character_>45} from Sec.~\ref{Sec:theta>45}. In fact, depending on the energy plane with $s=\sgn(\varepsilon-\varepsilon^0)$, one vortex at $\tilde\bK^s$ in the vortex pair moves to $s\infty$ while the one left at $\tilde\bK^{-s}$ approaches $\tilde\bK^*$. This topological transition at the critical angle $\theta=\pi/4$ qualitatively changes the vortex profile and can be intuitively understood since the vortex-vortex pair cannot annihilate, in contrast to the vortex-antivortex pair when $\theta<\pi/4$. 

\subsubsection{\texorpdfstring{$\theta<\pi/4$ case}{<45}}\label{Sec:theta<45}

As aforementioned, Eqs.~\eqref{eq:ky^s}\eqref{eq:barp12^s}\eqref{eq:borderline_diff} for vortex solutions still hold when $\theta<\pi/4$ as long as Eq.~\eqref{eq:Delta_barp1} is positive. This conditions leads to two critical energy planes
\begin{equation}\label{eq:epsilon_vt}
    \varepsilon_\mathrm{vt}^t = \frac{-\gamma_0\gamma d_3\cos\theta+t\gamma\sqrt{\cos{2\theta}(\gamma_0^2-\gamma^2)(d_3^2+Q)}}{Q}
\end{equation}
with $t=\pm1$ and $Q$ is given by Eq.~\eqref{eq:Q_def}.
Note that here we deliberately use superscript $t$ in order to distinguish from $s=\pm1$ mainly reserved for pair of monopoles or pair of vortices. Now, Eq.~\eqref{eq:borderline_diff} implies that $|\alpha'|>g>0$, leading to $\alpha'+s \,g\gtrless0$ when $\sgn(\alpha')=\pm1$ for both  $s=1$ and $s=-1$ and hence
\begin{equation}
    -\sgn(\tilde K_y^s - \tilde K_y^0)=\sgn[\alpha'(\varepsilon)]=\sgn[\varepsilon-\varepsilon^0]
\end{equation}
where the last equality takes into account Eq.~\eqref{eq:alpha_def} and the special energy plane $\varepsilon^0$ given by Eq.~\eqref{eq:epsilon^0_def}. Further, in the vicinity of the cores $\tilde K_y^s$ given by Eq.~\eqref{eq:ky^s}, Eq.~\eqref{eq:barp12^s} now yields the spin texture up to the leading order 
\begin{equation}
    \bar\bp_{12}^s=\left( -s\sqrt{\Delta}\delta\tilde k_y,2\sec2\theta\,(\alpha'-s\,g)k_x \right)\sim(-s\delta\tilde k_y,\sgn(\alpha)k_x),
\end{equation}
where the last expression leaves all but the sign information.
It gives a \textit{vortex-antivortex pair} as shown in Fig.~\ref{Fig:theta<45}, which is characterized by
\begin{equation}\label{eq:vt_character_<45}
    \mathtt{v}=s\chi\sgn(\alpha),\mathtt{h}=\sgn(\alpha)\frac{\pi}{2}.
\end{equation}
Directly related to $\sgn(\alpha)$, we also notice that the special energy $\varepsilon^0$ satisfies
\begin{equation}\label{eq:epsilonvt0_relation}
    \varepsilon_\mathrm{vt}^-\leq\varepsilon^0\leq\varepsilon_\mathrm{vt}^+,
\end{equation}
which can be seen with the aid of Eq.~\eqref{eq:Delta_barp1}, i.e., $\Delta(\varepsilon^0)=-\frac{1}{4}\cos{2\theta}(\gamma_0^2-\gamma^2)({\varepsilon^0}^2+\gamma^2)\leq0$. Note that the equality is realized when $\theta=\pi/4$.
In the case when $\theta$ approaches the critical angle $\pi/4$ from the above, the two vortices with the same vorticity cannot annihilate as mentioned in Sec.~\ref{Sec:theta=45}. In contrast, in the present case such a pair of vortex and antivortex can annihilate, which occurs at $(\varepsilon_\mathrm{vt}^t,\tilde\bk^t)$ with
\begin{equation}\label{eq:tildebk^t}
    \tilde\bk^t=(0,\frac{\sin\theta(\gamma_0\varepsilon_\mathrm{vt}^t+2\gamma d_3\cos\theta)}{\gamma\cos2\theta}).
\end{equation}
Last but not the least, we can connect the present case with the critical case $\theta=\pi/4$ in Sec.~\ref{Sec:theta=45}. In fact, when $\theta$ starts to decrease from $\pi/4$, both of $\varepsilon_\mathrm{vt}^\pm$ deviate from $\varepsilon^0$ in the manner of Eq.~\eqref{eq:epsilonvt0_relation}. In the energy region $\sgn(\varepsilon-\varepsilon_\mathrm{vt}^{t=s})=s$, the single vortex at $\tilde\bK^*$ will become the new vortex at $\tilde\bK^s$ with vorticity $\mathtt{v}=\chi$; the other antivortex is drawn from $-s\infty$ to the new finite position $\tilde\bK^{-s}$.

Combining these results, we reach the following picture that is largely distinct from the $\theta>\pi/4$ case or the similar pure $\gamma_1$ case. 
\begin{itemize}
    \item When $\theta<\pi/4$, one has a vortex-antivortex pair for $\varepsilon$-planes satisfying $\varepsilon\gtrless\varepsilon_\mathrm{vt}^\pm$. As shown in Fig.~\ref{Fig:theta<45}, the whole pair lies along the $\tilde k_y$-axis either below ($\varepsilon>\varepsilon_\mathrm{vt}^+$) or above ($\varepsilon<\varepsilon_\mathrm{vt}^-$) the spin-unidirectional line, which becomes no longer a borderline inside a pair. Moving to the inner region between the two critical $\varepsilon_\mathrm{vt}^\pm$ planes, the pair annihilates as a distinct topological transition.
\end{itemize}

\section{Experimental estimation}\label{SecSM:estimation}
Below, we discuss the estimation towards realistic experimental detection. As mentioned in the main text, we introduce the characteristic energy $\varepsilon'$, momentum $k'$, and relaxation strength $\gamma'$. We use the physically plausible assumption that $\gamma_\nu,\gamma_0\pm\gamma_i$ are roughly all of the order of magnitude of $\gamma'$.
We typically consider the case when $\varepsilon'$, i.e.,  either the energy $\varepsilon$ or the mass $m$, is not too close to zero, otherwise the situation becomes degenerate or too simple. From topological insulator surface state pumping measurements, one can estimate the spin relaxation time to be of the order $3\textrm{-}12\mathrm{ps}$ \cite{Cacho2015,Iyer2018}. This scale typically leads to $\gamma'\sim 0.4\textrm{-}2\mathrm{meV}$ and could be even larger in the presence of magnetization that enhances the magnetic impurity scattering, although it is usually not larger compared to $\varepsilon'$ that is often of several tens of milli-electronvolts. For the topological insulator surface state with magnetic dopants, experimentally it has been seen that the disorder can lead to considerable effects: the low quantization accuracy and observation temperature of the quantum anomalous Hall effect and other new phases related to magnetism are considered to be strongly affected by the magnetic disorder and the induced relaxation $\gamma'$ that, among other effects, gives the considerable density of states even within the gap \cite{Tokura2019,Wang2021}. %For instance, tunable and intense disorder at the nanoscale and hence large enough $\gamma'$ are generally confirmed for chromium-doped and other codoped topological insulators \cite{Lee2015,Liu2020,Tokura2019,Wang2021}. Such a situation is therefore very suitable for the experimental realization of the NH topological spin textures.

Under the above specifications, from Eq.~\eqref{eq:p_nu_gamma3_maintext} we have
\begin{equation}\label{eq:p_nu_gamma3_est}
\begin{split}
    \bp_{12}&\sim \gamma'\varepsilon'\bd_{12} \\
    p_{3}&\sim \gamma'(\varepsilon'^2-k^2).
\end{split}
\end{equation}
On the other hand, from Eq.~\eqref{eq:p_nu_gamma1_maintext} we have
\begin{equation}\label{eq:p_nu_gamma1_est}
\begin{split}
    \bp_{12}&\sim \gamma'\varepsilon'\bd_{12} + (\varepsilon'^2-k^2) \bgamma_1\\
    p_{3}&\sim \gamma'\varepsilon'm.
\end{split}
\end{equation}
These expressions imply that $\gamma_i$ and hence $\gamma'$ do \textit{not} directly affect the fine structure of the spin texture; instead, larger $\gamma'$ can enhance the signal strength as physically expected. %This is a very beneficial feature for realistic SARPES measurements, since the typical $\bk$-space scales of those fine textures will not become too small when the relaxation $\gamma_i$ are small, which as well partially reflects the topological robustness of those textures. 

In fact, the typical $\bk$-space scales of those fine textures are more related to $\varepsilon'$ and the Fermi velocity $v$. More specifically, from Eq.~\eqref{eq:p_nu_gamma3_est}, we immediately see that, in the $\gamma_3$ case, the typical skyrmion size, measured from the core to the radius where spins point in-plane, is given by $k'=\varepsilon'/v$. On the other hand for the $\gamma_1$ case, related to Eq.~\eqref{eq:p_nu_gamma1_est}, we actually can directly use the $\bK^\pm=\left(0,\frac{1}{\gamma_1}\left[\chi\gamma_0\varepsilon\pm\sqrt{(\gamma_0^2-\gamma_1^2)(\varepsilon^2+\gamma_1^2)+\gamma_1^2m^2}\right]\right)$ expression from the main text. Half of the meron pair separation measured between two centers is just given by $(\bK^+-\bK^-)/2$, which again leads to $k'=\varepsilon'/v$.

\end{document}